\documentclass[aps,nofootinbib,noshowpacs,showkeys,preprintnumbers,amsmath,amssymb]{revtex4}
\def\bes{\begin{subequations}}
\def\ees{\end{subequations}}
\def\be{\begin{equation}}
\def\ee{\end{equation}}
\def\bea{\begin{eqnarray}}
\def\eea{\end{eqnarray}}
\def\ba{\begin{eqnarray}}
\def\ea{\end{eqnarray}}
\def\bear{\begin{array}}
\def\eear{\end{array}}
\def\p1sl{\displaystyle{\not}p_1}
\def\p2sl{\displaystyle{\not}p_2}

\newcommand{\K}{{\widetilde {\cal K}}}

\newcommand{\blam}{{\overline \lambda}}
\usepackage{mathtools}
\newcommand{\fsl}[1]{\ensuremath{\mathrlap{\!\not{\phantom{#1}}}#1}}
\usepackage{graphics}
\usepackage{graphicx}
\usepackage{dcolumn}
\usepackage{bm}
\usepackage{epsfig}
\usepackage{graphicx}
\usepackage{multirow}
\usepackage{dcolumn}
\usepackage{graphicx,epsfig}%
\usepackage{color}

\begin{document}
\preprint{USM-TH-361}

\title{Sensitivity bounds on heavy neutrino mixing $|U_{\mu N}|^2$ and  $|U_{\tau N}|^2$ from LHCb upgrade}
\author{Gorazd Cveti\v{c}$^1$}
\email{gorazd.cvetic@usm.cl}
\author{C.~S.~Kim$^2$}
\email{cskim@yonsei.ac.kr}
\affiliation{$^1$
Department of Physics, Universidad T\'ecnica Federico Santa Mar\'ia, Valpara\'iso, Chile\\
$^3$Department of Physics and IPAP, Yonsei University, Seoul 120-749, Korea}

\date{\today}

\begin{abstract}
\noindent
Decays of heavy pseudoscalar mesons $B$, $B_c$, $B_s$ and $D_s$ at LHCb upgrade are considered, which produce either two equal sign muons or taus. In addition, we consider the analogous decays with opposite sign muons or taus. All these decays are considered to be mediated by a heavy on-shell neutrino $N$. Such decays of $B$ mesons, if not detected, will give in general stringent upper bounds on the heavy-light mixing parameter $|U_{\mu N}|^2$ as a function of the neutrino mass $M_N \sim 1$ GeV, principally due to the large expected number of produced mesons $B$. While some of the decays of the other mentioned mesons are attractive due to a weaker CKM-suppression, the expected produced number of such mesons is significantly smaller that that of $B$'s; therefore, the sensitivity bounds from such decays are in general comparable or less restrictive. When $\tau$ pairs are produced, only two types of such decays are significant: \textcolor{black}{$B^{\pm}, B_{c}^{\pm} \to \tau^{\pm} \tau^{\pm} \pi^{\mp}$ (and $\tau^{\pm} \tau^{\mp} \pi^{\pm}$),} giving us stringent upper bounds on $|U_{\tau N}|^2$; the other decays with a pair of $\tau$, such as \textcolor{black}{$B^0 \to D^{(*)-} \tau^+ \tau^+ \pi^-$ (and $D^{(*)-} \tau^+ \tau^- \pi^+$),} are prohibited or very suppressed by kinematics.
\end{abstract}
\keywords{rare meson decays; heavy neutrino; mixing parameters of heavy neutrino}

\maketitle


\section{Introduction}
\label{intr}

Various theories which can explain the masses of the three light neutrinos suggest the existence of heavy neutrinos, often referred to as sterile neutrinos. The neutrinos in such theories are of the Majorana type, i.e., they are their own antiparticles. As a consequence, they can produce, apart from lepton number conserving (LNC) processes, lepton number violating (LNV) processes through on-shell mediation. Dirac neutrinos can participate only in LNC processes. However, at present there are many open questions in the neutrino sector, among them: (a) What is the nature of the neutrinos, i.e., are they Majorana or Dirac particles? (b) How many heavy neutrinos exist and what are the values of their masses? (c) What are the values of their heavy-light mixing parameters $U_{\ell N}$ (where $\ell = e, \mu, \tau$)?

The Majorana nature of the neutrinos can be established by detection of LNV processes, such as neutrinosless double beta decay ($0\nu\beta\beta$) \cite{0nubb}, by specific scattering processes \cite{scatt1,scatt2}, by LNV rare decays of hadrons (usually mesons) \cite{mesdec1,Atre,SKetal,mesdecOur,symm,mesdecnew,mesdecQuint,mesdecShrock,mestaudec}, $\tau$'s \cite{mestaudec,taudec,CPtau} and heavy gauge bosons $W$ \cite{Wdec} and $Z$ \cite{Zdec}. In most of these processes, the analogous LNC channels also occur (cf.~in particular \cite{mesdecShrock}); the correct identification of such processes may sometimes have more severe background problems though. 

In \textcolor{black}{some} of these processes, the neutrino mass can also be determined or constrained. \textcolor{black}{If a considered pseudoscalar meson $M$ decays into invisible channel, $M \to$ invisible, then it is difficult or impossible to extract or constrain the mass $M_N$ of the involved neutrino(s) $N$ (via invisible width measurement), principally because we have competitive irreducible SM background $M \to \nu {\bar \nu} \nu {\bar \nu}$ \cite{BhaGao}\footnote{\textcolor{black}{Due to the helicity suppression, the SM background $M \to \nu {\bar \nu}$ is negligible \cite{BhaGao} in comparison.}} However, if a rare decay process of $M$ is mediated by an on-shell neutrino $N$ ($M_N \sim 1$ GeV) and, simultaneously, all or most of the final state particles are detectable, then the mass $M_N$ can be either determined ($M_N^2 = P_f^2$) or at least reasonably constrained. In this work we consider the rare decays $M \to N \ell \to \ell \ell \pi^{\pm}$ where $\ell=\mu$ or $\tau$, i.e., all the final state particles are charged and in principle detectable.} 

Furthermore, the phenomenon of oscillations \cite{neuosc} between the three light neutrinos has been observed \cite{oscobs}. If heavy neutrinos have almost degenerate mass, they can also oscillate \cite{oscheavy}. A somewhat related phenomenon is the resonant CP violation which can arise in scattering processes \cite{CPscatt}, rare meson decays \cite{symm,CKZCP1,CKZCP2,DCK} and rare $\tau$ decays \cite{CPtau}. There are several models with almost degenerate heavy neutrinos, among them the low-scale seesaw models \cite{lsseesaw} and the neutrino minimal standard model ($\nu$MSM) \cite{nuMSM1,nuMSM2}.

Various models with seesaw mechanism \cite{seesawLNV} and related models explain very low masses of the light neutrino sector, and contain heavy neutrinos with masses $M_N \gg 1$ TeV \cite{seesaw}, $M_N \sim 1$ TeV \cite{seesaw1TeV} and $M_N \sim 1$ GeV \cite{scatt2,nuMSM1,seesaw1GeV}. In the latter type of models, not only the masses are relatively low, but the heavy-light mixing parameters $|U_{\ell N}|^2$ are often less suppressed than in the earlier seesaw models \cite{seesaw}. Our analysis will be made with a view to such scenarios, i.e., $M_N \sim 1$ GeV and $|U_{\ell N}|^2 \sim 10^{-6}$-$10^{-4}$ (for $\ell=\mu, \tau$).

In this work we continue and extend the analysis of Ref.~\cite{SLCK} of the sensitivity bounds on the heavy-light mixing parameter $|U_{\mu N}|^2$ in the LHCb upgrade, where the rare lepton number violating (LNV) decays of $B$-mesons with two equal sign muons (LNV) were considered. Here we recalculate these rare $B$-decays, with the updated input parameters \cite{PDG2018}, and calculate also the analogous decays with opposite sign muons (LNC) when the neutrino $N$ is Dirac. \textcolor{black}{If the intermediate neutrino $N$ is Dirac, only LNC processes take place; on the other hand, if $N$ is Majorana, both LNC and LNV processes take place.}  In addition, we consider in this work the rare LNV and LNC decays of the pseudoscalar mesons $M =B_c$, $B_s$ and $D_s$, with two muons; and the rare decays of $M=B$ and $B_c$ with two taus. We consider that the decays are mediated by a heavy on-shell Majorana or Dirac neutrino $N$ with mass $M_N \sim 1$ GeV. The considered decays with muons involve the meson decay into vector $V$ or pseudoscalar meson $S$ and an off-shell $W^{*}$, where the latter gives a muon and and (on-shell) neutrino $N$:  $M \to V(S) \mu N$. The neutrino $N$ propagates from the primary vertex and decays at the secondary vertex within the detector, producing a muon and a pion:  $M \to V(S) \mu N \to V(S) \mu \mu \pi$. The considered decays are: (a) for $B$ mesons $B \to D^{*}(D) \mu \mu \pi$, and $B \to \mu \mu \pi$; (b) for $B_c$ mesons $B_c \to J/\Psi(\eta_c) \mu \mu \pi$, and  $B_c \to \mu \mu \pi$; (c) for $B_s$ mesons $B_s \to D_s^{*}(D_s) \mu \mu \pi$; (d) and for $D_s$ mesons $D_s \to \Phi(\eta) \mu \mu \pi$, and  $D_s \to \mu \mu \pi$. When the two charged leptons are $\tau$'s in the decays of the mentioned types, the kinematics allows only the consideration of two decays: $B \to \tau \tau \pi$ and $B_c \to \tau \tau \pi$.

\section{The formalism used}
\label{sec:form}

The analysis in Ref.~\cite{SLCK}, which we continue using here, was based on the formalism and expressions obtained in Ref.~\cite{CK1}. As in those references, we take the simplest representative scenario of heavy neutrinos, namely one heavy Majorana or Dirac neutrino $N$ (with mass $M_N \sim 1$-$10$ GeV), which has suppressed (heavy-light) mixing parameters $U_{\ell N}$ with the light flavor neutrinos $\nu_{\ell}$ ($\ell = e, \mu, \tau$)
\be
\nu_{\ell} = U_{\ell N} N + \sum_{k=1}^3 U_{\ell \nu_k} \nu_k \ .
\label{mixN}
\ee
Here, $\nu_k$ ($k=1,2,3$) are the three light neutrino mass eigenstates.

Here we summarize the expressions obtained in Ref.~\cite{CK1}.
The decay widths for $M \to V(S) \ell_1 N \to V(S) \ell_1 \ell_2 \pi$ are
\be
 \Gamma \left(M \to V(S) \ell_1 N \to V(S) \ell_1 \ell_2 \pi \right)
 =  \Gamma \left(M \to V(S) \ell_1 N \right) \frac{\Gamma(N \to \ell_2 \pi)}{\Gamma_N} \ .
\label{fact}
\ee
Here $\ell_j$ denote charged leptons; later we will have $\ell_1=\ell_2=\mu$ or $\ell_1=\ell_2=\tau$. The first factor, $\Gamma \left(M \to V(S) \ell_1 N \right)$, was calculated in Ref.~\cite{CK1} for the case of vector (V) and scalar (S) meson, and we refer to that reference for details. We note that the mass $M_N \sim 1$ GeV plays an important role. The second factor, $\Gamma(N \to \ell_2 \pi)$, is well known, and is also given explicitly in Refs.~\cite{CK1,SLCK}. The total decay width $\Gamma_N \equiv \Gamma(N \to {\rm all})$ was given and evaluated numerically as a function of $M_N$ in Ref.~\cite{CKZCP2} for the case of Majorana neutrinos, and in Ref.~\cite{symm} for the cases of Dirac and Majorana neutrinos.

As pointed out in \cite{SLCK}, the branching ratio of the process has to be multiplied by the probability $P_N$ of the intermediate $N$ neutrino to decay within the detector. This factor has to be considered with care, because it depends on the kinematic parameter $\beta_N''$which is the velocity of $N$ in the lab frame ($\Sigma''$). This $\beta_N''$ depends in a nontrivial way on the quantities $q^2$, ${\hat q}'$, and   ${\hat p}_1$, where $q$ is the four-momentum of the virtual $W^{*}$ ($\to \ell_1 N$), ${\hat q}'$ is its unitary direction vector in the $M$-rest frame ($\Sigma'$), and ${\hat p}_1$ is the direction of of the produced $\ell_1$ in the $W^{*}$-rest frame (i.e., $\ell_1 N$-pair rest frame, $\Sigma$)
\be
P_N = \left\{ 1 -
\exp \left[-
  \frac{L \Gamma_N}{\sqrt{ \left(E''_N(q^2;{\hat q}',{\hat p}_{1})/M_N \right)^2 - 1 }}
  \right] \right\}.
\label{PN}
\ee
Here, $E''_N$ is the energy of $N$ neutrino in the laboratory frame ($\Sigma''$), and $L$ is the effective length of the detector.\footnote{
The effective length $L$ in this sense is considered to be independent of the position in which the vertex of production of $N$ is situated, and independent of the direction in which the produced $N$ travels.}
  We refer to \cite{SLCK} for details. The true (effective) branching ratio ${\rm Br}_{\rm eff}$ of the process is then evaluated by integrating the corresponding differential decay rates multiplied by the decay probability $P_N$ and divided by $\Gamma_M$ 
\bea
{\rm Br}_{\rm eff}(M \to V(S) \ell_1 N \to V(S) \ell_1 \ell_2 \pi) & = &
    \frac{1}{\Gamma_M} \int d q^2 \int d \Omega_{{\hat q}'} \int d \Omega_{{\hat p}_1}
    \frac{d \Gamma(M \to V(S) \ell_1 N)}{ d q^2 d \Omega_{{\hat q}'}  d \Omega_{{\hat p}_1}} \frac{ \Gamma(N \to \ell_2 \pi) }{\Gamma_N}
\nonumber\\
&& \times \left\{ 1 - \exp \left[- \frac{L \Gamma_N}{\sqrt{ \left(E''_N(q^2;{\hat q}',{\hat p}_{1})/M_N \right)^2 - 1 }} \right] \right\}.
\label{Breff}
\eea
The  values of the total decay widths $\Gamma_M$ were taken from \cite{PDG2018}: $\Gamma_{B^+} = 4.018 \times 10^{-13}$ GeV; $\Gamma_{B^0} = 4.330 \times 10^{-13}$ GeV; $\Gamma_{B_c} = 1.298 \times 10^{-12}$ GeV; $\Gamma_{B_s} = 4.326 \times 10^{-13}$ GeV; $\Gamma_{D_s} = 1.306 \times 10^{-12}$ GeV.

The decay widths $\Gamma_N \equiv \Gamma(N \to {\rm all})$ of $N$ Majorana and Dirac neutrinos are obtained using the expressions of Ref.~\cite{CKZCP2} (Appendix B and Fig.~2 there); cf.~also \cite{symm} (Appendix A.3 and Fig.~2 there).
They have the form
\begin{equation}
  \Gamma_{N} =  \K \; \frac{G_F^2 M_{N}^5}{96 \pi^3},
\label{GNwidth}
\end{equation}
where the factor $\K$ in Eq.~(\ref{GNwidth}) has all the dependence on the heavy neutrino mixing factors
\begin{equation}
  \K = {\cal N}_{e N} \; |U_{e N}|^2 +
    {\cal N}_{\mu N} \; |U_{\mu N}|^2 + {\cal N}_{\tau N} \; |U_{\tau N}|^2  \ .
\label{calK}
\end{equation}
The (dimensionless) numbers ${\cal N}_{\ell N}$  are functions of the mass $M_N$, ${\cal N}_{\ell N}(M_N) \sim 1$-$10$. They are determined by the formulas given in Appendix B of Ref.~\cite{CKZCP2} which are based on the approach of Ref.~\cite{SKetal}. The results for ${\cal N}_{\ell N}(M_N)$ are presented, e.g., as a curve in Fig.~2 of Ref.~\cite{CKZCP2}.

When the produced meson is vector ($V$), the differential decay width $d \Gamma(M \to V \ell_1 N)/(d q^2 d \Omega_{{\hat q}'}  d \Omega_{{\hat p}_1})$ is complicated \cite{CK1}, due to both the vector nature of the produced meson $V$ and the nonzero mass of $N$ (and of $\ell_1$); but it does not depend on the electric charge of $\ell_1$. The available literature does not shed light unequivocally on the latter point, for which we refer to the Appendix.

If no mesons $V$ or $S$ are produced, the differential decay width for $M \to \ell_1 N$ is simpler, it depends only on the direction ${\hat p}_N'$ of the on-shell $N$ in the $M$-rest frame; and since $M$ is (pseudo)scalar, we have $d \Gamma(M \to \ell_1 N)/d \Omega_{{\hat p}_N'}= \Gamma(M \to \ell_1 N)/(4 \pi)$. Further, the decay probability $P_N$ also depends only on ${\hat p}_N'$. This then gives (for details, cf.~(\cite{CK1,SLCK})
\bea
    {\rm Br}_{\rm eff}(M^{\pm} \to \ell_1^{\pm} N \to \ell_1^{\pm} \ell_2^{\pm} \pi^{\mp}) & = &
    \frac{1}{M_M} \frac{1}{4 \pi}
    \int d \Omega_{{\hat p}'_N} \Gamma(M^{\pm} \to \ell_1^{\pm} N) \frac{ \Gamma(N \to \ell_2^{\pm} \pi^{\mp}) }{\Gamma_N}
\nonumber\\
&&\times
\left\{ 1 - \exp \left[- \frac{L \Gamma_N}{\sqrt{ \left(E''_N({\hat p}'_N)/M_N \right)^2 - 1 }} \right] \right\}.
\label{BreffnoVS}
\eea

\section{Numerical results for the achievable upper bound limits of $|U_{\ell N}|^2$ ($\ell = \mu, \tau$) at LHCb upgrade}
\label{sec:num}

\begin{figure}[htb] 
\begin{minipage}[b]{.49\linewidth}
\includegraphics[width=85mm,height=50mm]{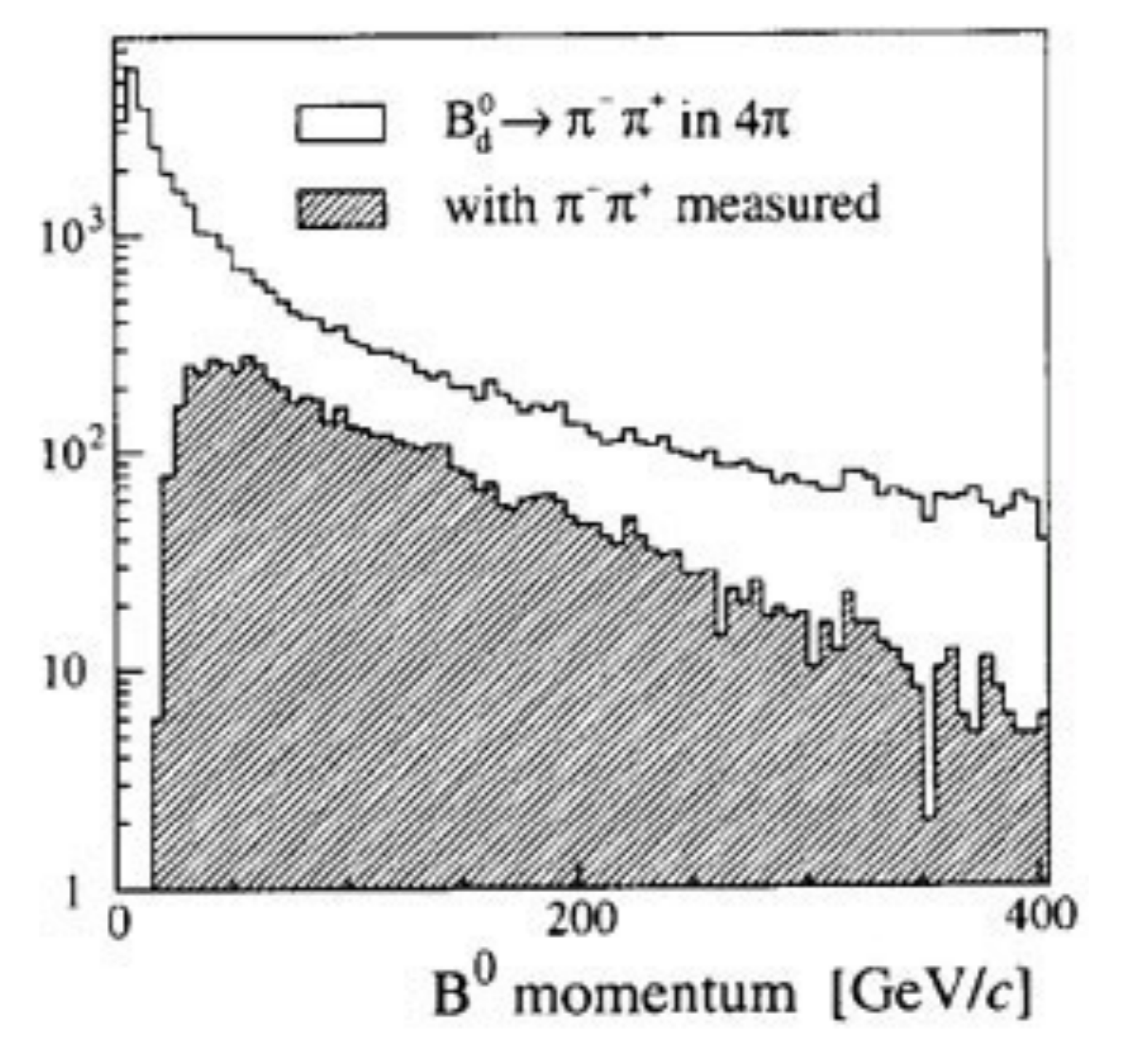}
\end{minipage}
\begin{minipage}[b]{.49\linewidth}
\includegraphics[width=75mm,height=47mm]{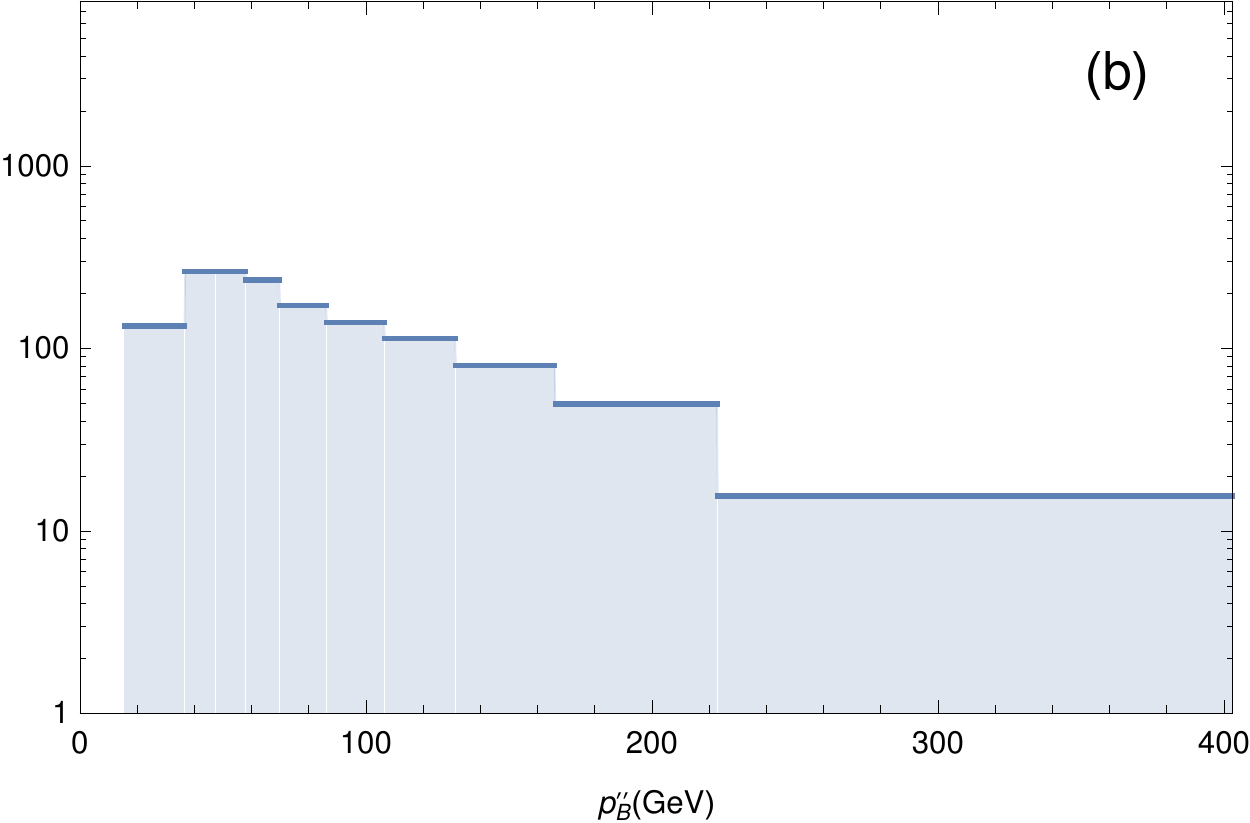}
\end{minipage} \vspace{12pt}
\caption{\footnotesize (a) The lab momentum  ($|{\vec p}_B''|$) distribution of the produced $B^0$ mesons in LHCb (the shaded figure); (b) partition of the left-hand shaded curve in ten bins of equal weight. The apparent bin with the largest height are in fact two bins (with almost equal height).}
\label{FigBdist}
\end{figure}

We will consider the mentioned rare decays of $M=B, B_s, B_c$ and $D_s$ at LHCb upgrade. We take into account that the produced $B$ mesons have a specific distribution of the momentum $(|{\vec p}_B|)_{\rm lab} \equiv |{\vec p}_B''|$ in laboratory ($\Sigma''$) frame, as given in Fig.~\ref{FigBdist}(a).\footnote{
  We thank Sheldon L.~Stone (LHCb Collaboration) for providing us with this distribution.}
In practice, we separated this distribution in ten bins with equal weight (i.e., equal number of events), cf.~Fig.~\ref{FigBdist}(b), and calculated the results by averaging over these ten bins. We used the same distribution also when $M=B_s$ and $M=B_c$. For the decays with $M=D_s$, on the other hand, we took $|{\vec p}_{D_s}''|=50$ GeV as a representative case, produced in LHCb by decays of $B_s$ (and ${\bar B}_s$) mesons.

Further arithmetic averaging is made in the $B$-decays by averaging over the modes with $B^{\pm}$ on the one hand and $B^{0}$ and ${\bar B}^{0}$ on the other hand, because the total decay widths are somewhat different in the two cases ($\Gamma_B=4.018 \times 10^{-13}$ GeV, $4.330 \times 10^{-13}$ GeV, respectively). The (partial) decay widths for the decays $M \to V \ell_1 N$ with $\ell_1^+$ and $\ell_1^-$ are the same, as mentioned in the previous Section and explained in the Appendix.  

The effective detector length $L$ in LHCb could be considered to be approximately the length of the Vertex Locator (VELO) which is about $1$ m \cite{VELO}. However, the length can be extended beyond the locator, to about $2.3$ m \cite{Sheldon}; we will take $L=2.3$ m.

For the values of the masses and CKM matrix elements we used the central values from \cite{PDG2018}. Similarly, we used the values of the decay constants (for the annihilation-type decays) $f_B=0.1871$ GeV \cite{PDG2018}, $f_{B_c}=0.322$ GeV \cite{CKWW} and $f_{D_s}=0.248$ GeV \cite{CKWW} (the average in Ref.~\cite{PDG2018} is practically the same, $f_{D_s}=0.249$ GeV). As mentioned in the previous Section, the total decay widths $\Gamma_N$ of $N$ Majorana or Dirac neutrinos are obtained using the expressions of Ref.~\cite{CKZCP2} (Appendix B there), based on the approach of Ref.~\cite{SKetal}.

Further, in our analysis for the rare decays with two muons in the final state we considered only the mixing elements $|U_{\mu N}|^2$ as nonzero; for those with two taus in the final state we considered only $|U_{\tau N}|^2$ as nonzero. If, in addition, other mixing elements were nonzero, this would increase the total decay width $\Gamma_N$, cf.~Eqs.~(\ref{GNwidth})-(\ref{calK}), and would, as a consequence, decrease the effective branching ratio ${\rm Br}_{\rm eff}$ of the considered rare decays and make the upper bound on the mixing element less stringent (higher).

The form factors used for the decays $B \to D \ell_1 N$ are from \cite{CLN,Belle15} ($F_1$) and \cite{CK1} ($F_0$); for the decays $B \to D^{*} \ell_1 N$ are from \cite{Belle2} ($V, A_1, A_2$) and \cite{CK1,SLCK} ($A_0$); for $B_s \to D_s^{(*)} \ell_1 N$ from \cite{Fanetal}; for $B_c \to J/\Psi (\eta_c) \ell_1 N$ from \cite{Wangetal}; for $D_s \to \eta \ell_1 N$ from \cite{DM} and for  $D_s \to \phi \ell_1 N$ from \cite{WSh}.

The effective branching ratios depend on the number of produced mesons $M$ whose rare decays we are considering. At the LHCb upgrade, the projected numbers are: $N_B=4.8 \times 10^{12}$; $N_{B_s}=5.76 \times 10^{11}$; $N_{B_c}=2.4 \times 10^{10}$ \cite{Sheldon}. The mesons $D_s$ are mainly produced by decays of $B_s$ mesons, with decay branching ratio around $0.9$; therefore we take $N_{D_s} = 0.9 N_{B_s} =5.18 \times 10^{11}$.

The sensitivity upper bounds on the heavy-light mixing parameters $|U_{\mu N}|^2$ and $|U_{\tau N}|^2$ at the 95\% confidence limit are then obtained by requiring $N_{\rm events}=3.09$ \cite{FC}. For example, for rare $B$ decays we require ${\rm Br}_{\rm eff} = 3.09/(4.8 \times 10^{12})$, and obtain the corresponding upper bounds on  $|U_{\mu N}|^2$ from the absence of rare decays producing two muons, and on $|U_{\tau N}|^2$ when the two leptons are taus.\footnote{\textcolor{black}{The 95\% confidence level on the upper bound refers to zero signal events and zero known background events. In the case of LNC events, the assumption of zero known (SM) background events is probably not realistic.}}

It turns out that when the two charged leptons are muons, all mentioned decays of $B$, $B_s$, $B_c$ and even $D_s$ are kinematically allowed. \textcolor{black}{The LNV versions of such decays are}
\bes
\label{decMMpmumuPi}
\bea
B^0 & \to & D^{(*)-} \mu^+ N \to  D^{(*)-} \mu^+ \mu^+ \pi^-, \qquad
B^+  \to  D^{(*)0} \mu^+ N \to  D^{(*)0} \mu^+ \mu^+ \pi^-,
\label{decB}
\\
B_s^0 &\to&  D_s^{(*)-} \mu^+ N \to  D_s^{(*)-} \mu^+ \mu^+ \pi^-,
\label{decBs}
\\
B_c^+ & \to & J/\Psi (\eta_c) \mu^+ N \to J/\Psi (\eta_c) \mu^+ \mu^+ \pi^-,
\label{decBc}
\\
D_s^+ & \to & \phi (\eta) \mu^+ N \to \phi (\eta) \mu^+ \mu^+ \pi^-,
\label{decDs}
\eea
\ees
\textcolor{black}{and their charge-conjugate versions.}
The decays of the above mesons are possible also when no vector (or pseudoscalar) mesons are produced, i.e., when the decays are of the annihilation-type. This is the case for the decays of charged $M =B$, $B_c$ and $D_s$
\be
M^{\pm} \to \mu^{\pm} N \to \mu^{\pm} \mu^{\pm} \pi^{\mp}, \quad (M^{\pm}=B^{\pm}, B_c^{\pm}, D_s^{\pm}).
\label{decMmumuPi}
\ee
When the two produced charged leptons are $\tau$'s, the decays of the type (\ref{decMMpmumuPi}) are kinematically not possible; the only decays with $\tau$'s that are kinematically possible are
\be
B^{\pm}, B^{\pm}_{c} \to \tau^{\pm} N \to  \tau^{\pm}   \tau^{\pm}  \pi^{\mp}.
\label{BtautauPi}
\ee
The mentioned decays, Eqs.~(\ref{decMMpmumuPi})-(\ref{BtautauPi}), are first considered to be LNV, i.e., the produced lepton pair is of equal sign ($\mu^{\pm} \mu^{\pm}$ or $\tau^{\pm} \tau^{\pm}$). In such a case, the neutrino $N$ is considered to be Majorana. The analogous LNC decays are the decays with opposite sign lepton pairs, \textcolor{black}{i.e.,  $\mu^{\pm} \mu^{\pm} \pi^{\mp}$ gets replaced by $\mu^{\pm} \mu^{\mp} \pi^{\pm}$, and $\tau^{\pm} \tau^{\pm} \pi^{\mp}$ by $\tau^{\pm} \tau^{\mp} \pi^{\pm}$.} The decay rates for the analogous LNC decays are equal to those of LNV decays \cite{CK1} when $N$ is Majorana neutrino; in such a case, such LNC decays give identical sensitivity limits for the mixing parameters ($|U_{\mu N}|^2$; $|U_{\tau N}|^2$) as the corresponding LNV decays. On the other hand, if $N$ is Dirac neutrino, only LNC decays are possible, and the total decay width $\Gamma_N$ of Dirac neutrino is smaller than that of the corresponding Majorana neutrino (with the same mass and the same mixing parameter values). This is so because Dirac neutrinos have a significantly smaller number of decay channels than the Majorana neutrinos, resulting in about 40 per cent smaller $\Gamma_N$, i.e., the coefficients ${\cal N}_{\ell N}$ in Eq.~(\ref{calK}) are by about 40 per cent smaller than in the Majorana case (cf.~Fig.~2 in Ref.~\cite{symm}). Smaller $\Gamma_N$ implies that the (weakly) $\Gamma_N$-dependent part of the integrand for the effective branching ratios ${\rm Br}_{\rm eff}$ in Eqs.~(\ref{Breff}) and (\ref{BreffnoVS})
\bea
\label{GNpart}
\frac{1}{\Gamma_N} \times
\left\{ 1 - \exp \left[- \frac{L \Gamma_N}{\sqrt{ \left(E''_N/M_N \right)^2 - 1 }} \right]
\right\} &=&
\frac{L}{\sqrt{ \left(E''_N/M_N \right)^2 - 1 }} - \frac{1}{2} \frac{L^2 \Gamma_N}{\left[ \left(E''_N/M_N \right)^2 - 1 \right]}
+ \ldots
\eea
becomes larger, and thus ${\rm Br}_{\rm eff}$ becomes larger. This implies that in the case of Dirac $N$ the sensitivity limits (upper bounds) on the mixing parameters become more restrictive (smaller). This effect is expected to be appreciable only when $\Gamma_N$ is large, i.e., when $M_N$ is large.

\begin{figure}[htb] 
\begin{minipage}[b]{.49\linewidth}
\includegraphics[width=85mm,height=50mm]{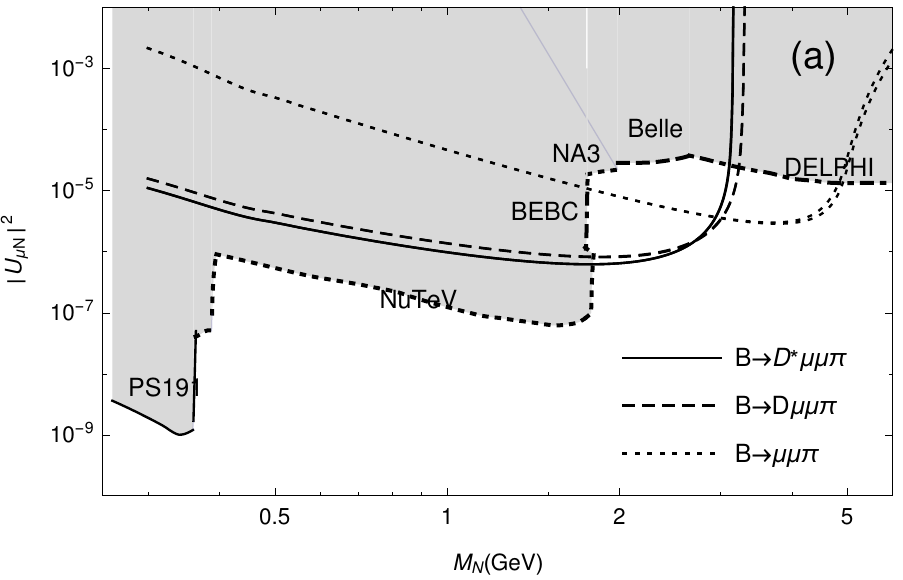}
\end{minipage}
\begin{minipage}[b]{.49\linewidth}
\includegraphics[width=75mm,height=47mm]{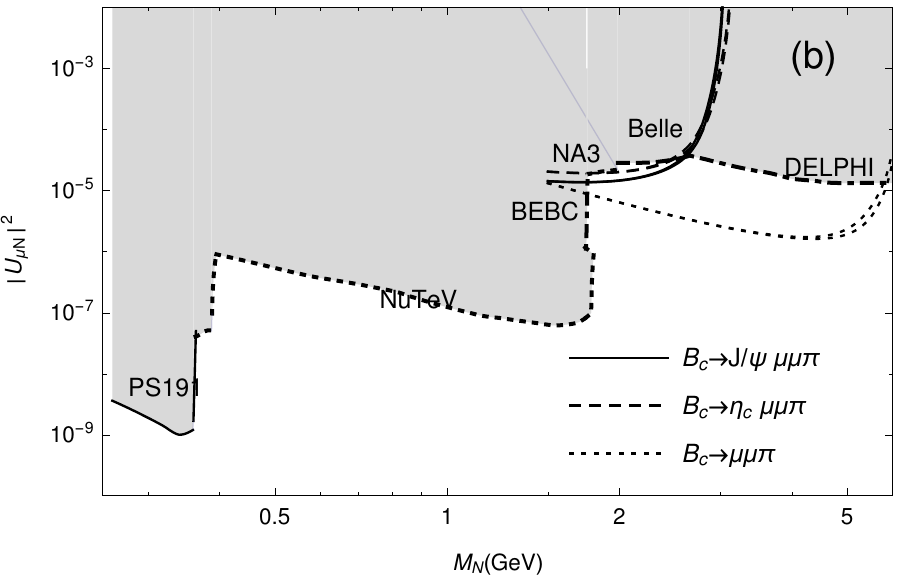}
\end{minipage}
\begin{minipage}[b]{.49\linewidth}
\includegraphics[width=85mm,height=50mm]{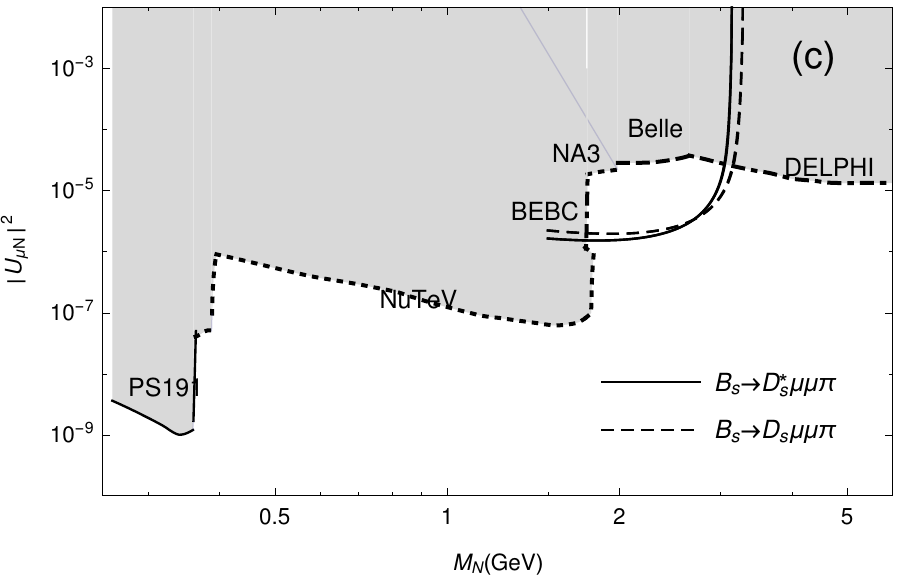}
\end{minipage}
\begin{minipage}[b]{.49\linewidth}
\includegraphics[width=75mm,height=47mm]{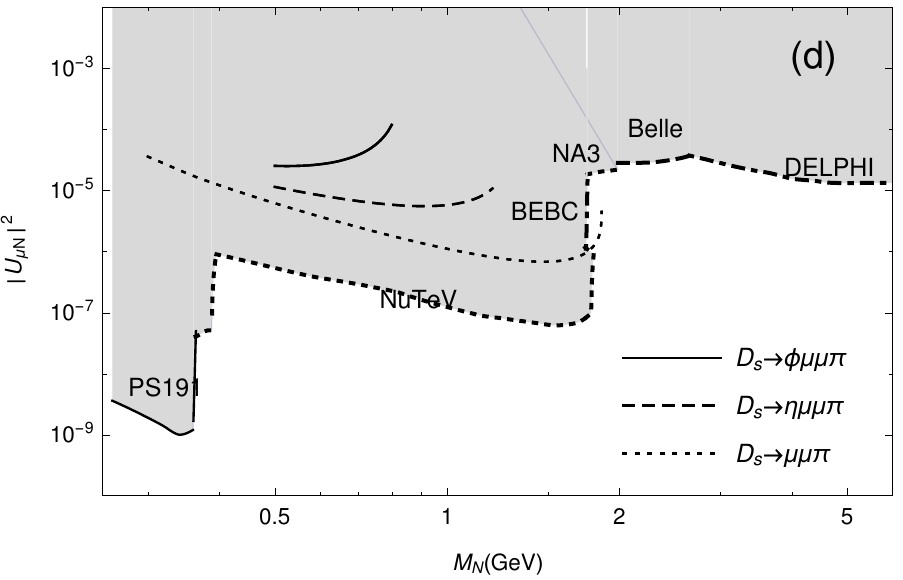}
\end{minipage}
\caption{Sensitivity limits for $|U_{\mu N}|^2$ from LHCb upgrade from LNV and LNC decays of $B$, $B_c$, $B_s$ and $D_s$ mesons with two equal sign or opposite sign muons in the final state. \textcolor{black}{The mass $M_N$ of the intermediate neutrino $N$ is kinematically restricted to be on-shell.} The See the text for futher details.}
\label{FigDecmu}
\end{figure}
The resulting sensitivity limits (upper bounds) on the heavy-light mixing parameter $|U_{\mu N}|^2$, as a function of mass $M_N$, for the decay processes involving muon pair are presented in Figs.~\ref{FigDecmu}(a)-(d). For each decay channel, we presented the limits from LNV decays (for $N$ Majorana) and the limits when $N$ is Dirac neutrino (LNC decays with $N$ Dirac); the latter limits are in general only slightly better (lower) and are clearly visible only in the case of $B_{(c)} \to \mu \mu \pi$ decays at large masses $M_N \approx 4.5$-$6.0$ GeV.
In these Figures we also included the present experimental upper bounds on this parameter from various experiments: the beam dump experiments PS191 \cite{PS191}, NuTeV \cite{NuTeV}, NA3 \cite{NA3} and BEBC \cite{BEBC}; Belle experiment \cite{BelleUB} which looked for the rare LNV decays $B \to X \ell N$ (and $N \to \ell \pi$); and DELPHI experiment \cite{DELPHI} which looked for rare $Z \nu N$ decays with subsequent $N$ decays. The present experimental LHCb upper bounds \cite{LHCbUB,PesShuUB}, from $B$ meson decays with two equal sign muons, are less restrictive than the combination of Belle and DELPHI bounds.

We can see that, within the kinematically allowed ranges, the present experimental sensitivity limits can be improved the most with $B \to D^{(*)} \mu \mu \pi$ decays for $1.75 \ {\rm GeV} < M_N < 2.95 \ {\rm GeV}$, and by $B_c \to \mu \mu \pi$ decays for $2.95 \ {\rm GeV} < M_N < 5.8 \ {\rm GeV}$. Decays $B_s \to D_s^{(*)} \mu \mu \pi$ give somewhat less improved bounds than $B \to D^{(*)} \mu \mu \pi$, in the mentioned mass interval $[1.75,2.95]$ GeV. On the other hand, the rare decays of $D_s$, Fig.~\ref{FigDecmu}(d), give bounds in the region of smaller $M_N < 1.75$ GeV (due to the smaller mass of $D_s$) where the present experimental bounds are more stringent (by NuTeV and PS191 experiments); the only exception is the small interval $1.75 \ {\rm GeV} < M_N < 1.85 \ {\rm GeV}$ where the decay $D_s \to \mu \mu \pi$ gives improved upper bounds.

The results of Fig.~\ref{FigDecmu}(a), for the rare $B$-meson decays, largely agree with the results obtained by us earlier in Ref.~\cite{SLCK}, the changes appearing mainly due to the updated values of the input parameters \cite{PDG2018}. We recall that the form factors $V$ and $A_j$ for these decays are taken from Ref.~\cite{Belle2} and $F_1$ from \cite{Belle15}, and there they were determined in such a way that the product of them with the CKM matrix element $|V_{cb}|$ is independent of the value of $|V_{cb}|$ (this does not apply to the form factor $F_0$ determined in \cite{SLCK}). Further, here we took into account that in the $B \to D^{*} \mu \mu \pi$ decay the results are independent of the sign of charge of the $\mu$ pair; while previously in \cite{SLCK} we assumed a (weak) dependence on this sign. However, this difference accounts for less than one per cent in the obtained upper bounds, cf.~the Appendix. Therefore, here we obtain for the sensitivity limits on $|U_{\mu N}|^2$ from $B \to D^{*} \mu \mu \pi$ practically the same values as in Ref.~\cite{SLCK}.
From $B \to D \mu \mu \pi$ we obtain here the values lower by about $2$-$4$ per cent than in Ref.~\cite{SLCK}, mostly due to the updated value of $|V_{cb}|$ in the form factor $F_0$. And from $B \to \mu \mu \pi$ the values of sensitivity limits  $|U_{\mu N}|^2$ are higher by about 4 per cent than in Ref.~\cite{SLCK}, mostly due to the updated value of $|V_{ub}|$ and of the decay constant $f_B$.

We point out that in Ref.~\cite{SLCK} no rare decays of $B_c$, $B_s$ and $D_s$ were considered. Here we can see that the decays of these mesons, with two muons in the final state, can be considered as complementary to the analogous $B$ meson decays.\footnote{The LNV decays $B_c \to J/\Psi \mu \mu \pi$, $\mu \mu \pi$, and $B_s \to D_s(K) \mu \mu \pi$ for LHCb were first considered in Ref.~\cite{mesdecQuint}, for the future LHC-run3 with assumed integrated luminosity of up to $50 \ {\rm fb}^{-1}$. For $B_c$ decays they obtained upper bounds $|U_{\mu N}|^2 \sim 10^{-5}$-$10^{-4}$ (we obtained $\sim 10^{-6}$-$10^{-5}$); for $B_s$ decays they obtained upper bounds somewhat less restrictive than the combined Belle and DELPHI bounds.}

\begin{figure}[htb]
\centering\includegraphics[width=90mm]{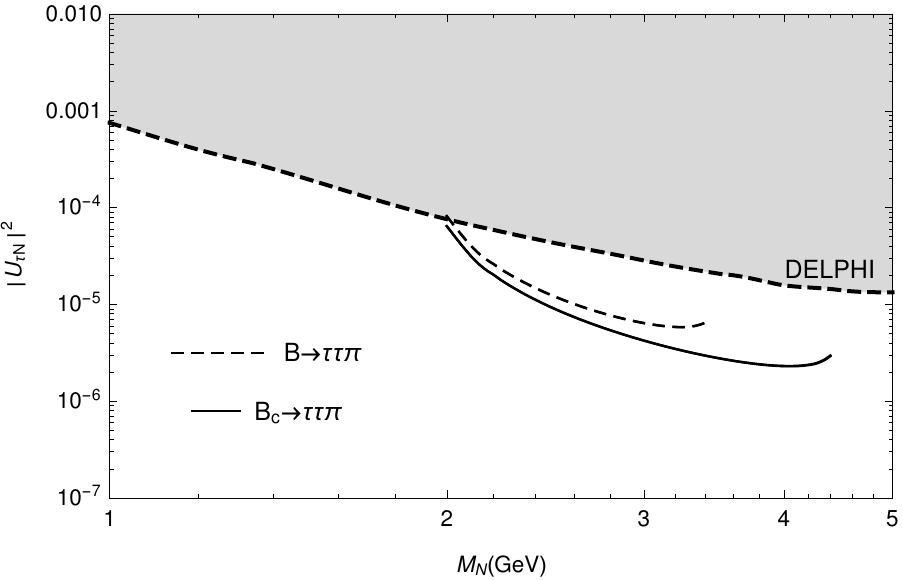}
\caption{Sensitivity limits for $|U_{\tau N}|^2$ at the LHCb upgrade from $B_{(c)}^{\pm} \to \tau^{\pm} N \to \tau^{\pm} \tau^{\pm} \pi^{\mp}$ decays. \textcolor{black}{The mass $M_N$ of the intermediate neutrino $N$ is kinematically restricted to be on-shell.}}
\label{FigDectau}
\end{figure}
When two $\tau$ leptons are in the final state, the only kinematically allowed decays are $B^{\pm}, B_c^{\pm} \to \tau^{\pm} N \to \tau^{\pm} \tau^{\pm} \pi^{\mp}$. Figure \ref{FigDectau} shows improved upper bounds on $|U_{\tau N}|^2$ from the decays $B_{(c)} \to \tau \tau \pi$. The improvement is better for $B_c$ decays, and it is in the larger kinematically allowed interval $2.0 \ {\rm GeV} < M_N < 4.4$ GeV. The sensitivity limits for the case of LNC decays with $N$ Dirac neutrino are only slightly lower, and the difference is practically invisible in the Figure. The present experimental bounds, in the presented mass range, are from DELPHI experiment \cite{DELPHI}. The new limits are important also in view of the fact that, at the moment, the available experimental limits on $|U_{\tau N}|^2$ are only from one experiment (DELPHI).  We point out that in Ref.~\cite{SLCK} no such decays were considered.

Some of the considered rare decays of $B_c$ and $B_s$ appear theoretically equally or more attractive than the corresponding rare decays of $B$ mesons. For example, the annihilation type decays $B_c  \to \ell N \to \ell \ell \pi$ are much less CKM-suppressed than the corresponding charged $B$-meson decays. Nonetheless, the obtained sensitivity limits on $|U_{\mu N}|^2$ are only a little better in the $B_c$ case. This is so because the expected number of the produced $B_c$ mesons in LHCb upgrade is significantly lower than the expected number of the produced $B$ mesons. Nonetheless, in the decays $B, B_c \to \tau N \to \tau \tau \pi$, the mesons $B_c$ give significantly better results; this is so  because the pair $\tau \tau$ is much heavier than $\mu \mu$, and thus the fact that $B_c$ has a higher mass than $B$ becomes important.

A similar argument applies to the decays of $D_s$: the sensitivity limits on $|U_{\mu N}|^2$ from $D_s \to \mu \mu \pi$ are less restrictive due to the suppressed expected number of produced $D_s$.

In summary, in this work we extended the analysis of our previous work \cite{SLCK} where the rare decays of $B$ mesons were considered. In the present analysis, we considered the rare decays of the mesons $M=B, B_c, B_s$ and $D_s$ in the future LHCb upgrade, where in the first vertex a vector (V) or pseudoscalar (S) meson is produced together with a charged lepton $\ell_1$ and an on-shell $\sim 1$ GeV Majorana or Dirac neutrino $N$: $M \to V(S) W^* \to V(S) \ell_1 N$. The produced $\sim 1$ GeV neutrino is assumed to travel and decay within the detector, $N \to \ell_2 \pi$. Further, also the annihilation-type rare decays were considered, i.e., those where no $V$ (or $S$) mesons are produced, $M^{\pm} \to W^{* \pm} \to \ell_1 N \to \ell_1 \ell_2 \pi$.  The two charged leptons were assumed to be equal sign muons or taus: $\ell_1 \ell_2 = \mu^{\pm} \mu^{\pm}$ or $\tau^{\pm} \tau^{\pm}$, i.e., the decays were assumed to be distinctly LNV. In addition, we considered also the case of Dirac neutrino $N$, where the analogous decays were LNC, i.e., $\ell_1 \ell_2 = \mu^{\pm} \mu^{\mp}$ or $\tau^{\pm} \tau^{\mp}$. It turned out that, in the case that the considered decays are not detected, the upper bounds on the mixing parameters $|U_{\mu N}|^2$ and $|U_{\tau N}|^2$ of $\sim 1$ GeV neutrino can be significantly improved.

\appendix

\section{Differential decay width for $M \to V \ell_1 N$}
\label{app:dG}

In this Appendix we clarify some aspects of the decays $M \to V \ell_1 N$ which appear ambiguous in parts of the literature.

\begin{figure}[htb]
\centering\includegraphics[width=90mm]{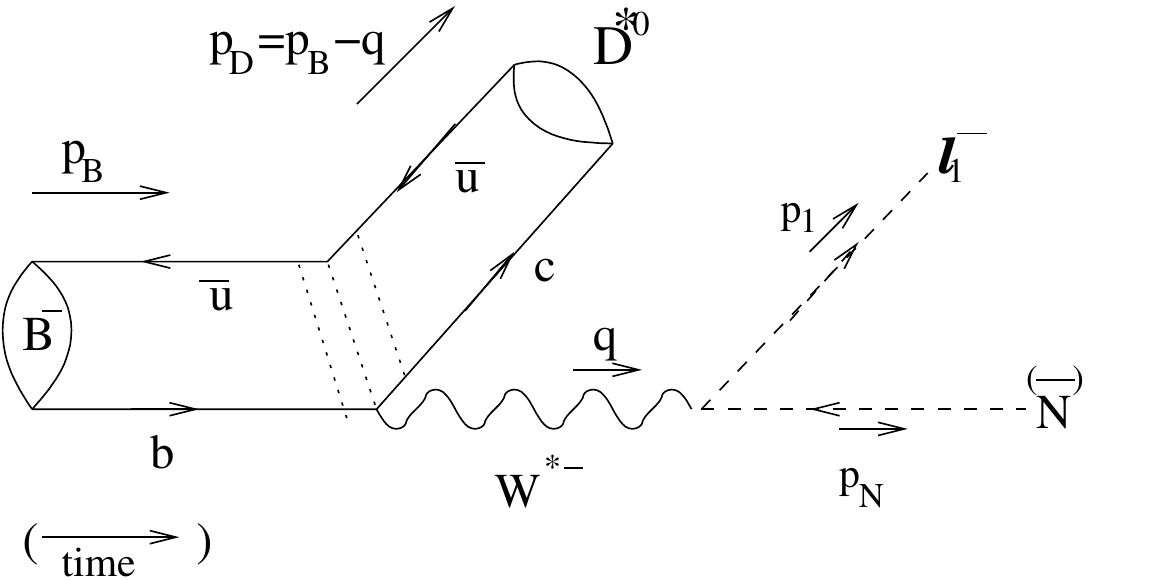}
\caption{Schematical representation of the decay $B^- \to D^{*0} \ell_1^- {\bar N}$. Other decays of the type $M \to V \ell_1 N$ are completely analogous: ${\bar B}^0 \to D^{*+ } \ell_1^- {\bar N}$;  ${\bar B}^0_s \to D_s^{*+ } \ell_1^- {\bar N}$;  $B_c^- \to J/\Psi \ell_1^- {\bar N}$; $D_s^- \to \phi \ell_1^- {\bar N}$.}
\label{FigBDstW}
\end{figure}
When the pseudoscalar $M$ decays into a vector particle $V$, charged lepton $\ell_1$, and heavy ($\sim 1$ GeV) neutrino $N$, the differential decay width is significantly more complicated than when a (pseudo)scalar $S$ is produced instead of $V$. Further, it is more complicated than the differential decay width when the neutrino and the charged leptons are (practically) massless. The decay $M \to V \ell_1 N$ is presented schematically in Fig.~\ref{FigBDstW}. The complexity is reflected in the structure of the hadronic matrix elements; in the specific case when $M=B$ (and $V=D^{*}$) we have
\bes
\label{Hmu}
\bea
H^{\mu}_{(\eta=-1)} &\equiv& \langle D^{*-}(p_D) | {\overline b} \gamma^{\mu} (1 - \gamma_5) c | B^{0}(p_B) \rangle =
\langle {\overline D^{*0}}(p_D) | {\overline b} \gamma^{\mu} (1 - \gamma_5)  c | B^{+}(p_B) \rangle,
\label{Hetam}
\\
H^{\mu}_{(\eta=+1)} &\equiv& \langle D^{*+}(p_D) | {\overline c} \gamma^{\mu} (1 - \gamma_5) b | {\overline B^{0}}(p_B) \rangle =
\langle D^{*0}(p_D) | {\overline c} \gamma^{\mu} (1 - \gamma_5) b | B^{-}(p_B) \rangle \ ,
\label{Hetap}
\eea
\ees
while in the cases of $M=B_s$ ($V=D_s^*$), $M=B_c$ ($V=J/\Psi$) and $M=D_s$ ($V=\phi$) the matrix elements $H^{\mu}_{\eta}$ are completely analogous. These matrix elements are expressed with form factors $V$ and $A_j$ ($j=0,1,2$) in the following way:
\bea
H^{\mu}_{\eta}
& = &
- i 2 \eta  \frac{\varepsilon^{\mu \nu \alpha \beta}}{(M_M+ M_V)} \epsilon^*_{\nu} (p_D)_{\alpha} (p_B)_{\beta} V(q^2) - \left[ (M_M+M_V) \epsilon^{* \mu} A_1(q^2)
  - \frac{\epsilon^* \cdot q}{(M_M+M_V)} (p_B+p_D)^{\mu} A_2(q^2) \right]
\nonumber\\
&& + 2 M_V \frac{\epsilon^* \cdot q}{q^2} q^{\mu} \left( A_3(q^2) - A_0(q^2) \right) \ ,
\label{FFBDst}
\eea
and $A_3(q^2)$ denotes
\be
A_3(q^2) = \frac{1}{2 M_V} \left[ (M_M+M_V) A_1(q^2) -
(M_M-M_V)
A_2(q^2) \right] .
\label{A3}
\ee
The first term in Eq.~(\ref{FFBDst}) has a factor $\eta = \pm 1$, which is obtained by the application of charge conjugation operation to the hadronic matrix elements; we have $\eta=+1$ when $\ell_1^-$ is produced, and $\eta=-1$ when $\ell_1^+$ is produced. More explicitly, the (reduced) decay amplitude for the process  $B \to D^{*} \ell_1 N$ is
\bes
\label{TBDstNl}
\bea
{\cal T}_{(\eta=-1)}  &=&  U^*_{\ell_1 N} V^*_{c b} \frac{G_F}{\sqrt{2}} \left[{\overline u}_{(N)}(p_N) \gamma_{\mu} (1 - \gamma_5) v_{(\ell_1)}(p_1) \right] H^{\mu}_{(\eta=-1)} \ ,
\label{TBDstNletm}
\\
{\cal T}_{(\eta=+1)}  &=&  U_{\ell_1 N} V_{c b} \frac{G_F}{\sqrt{2}} \left[{\overline u}_{(\ell_1)}(p_1) \gamma_{\mu} (1 - \gamma_5) v_{(N)}(p_N) \right] H^{\mu}_{(\eta=+1)} \ ,
\label{TBDstNletp}
\eea
\ees
When making the absolute square and summing over the final leptonic helicities leads to
\be
|{\cal T}|^2 = |U_{\ell_1 N}|^2 |V_{c b}|^2 \frac{G_F^2}{2} L^{\mu \nu} H_{\mu} H_{\nu}^* \ ,
\label{Tsq1}
\ee
where $L^{\mu \nu}$ is the lepton tensor
\bea
L^{\mu \nu} & = & 2 {\rm tr} \left[ \fsl{p}_N \gamma^{\mu} \fsl{p}_1 \gamma^{\nu} (1 + \eta \gamma_5) \right]
\nonumber\\
& = & 8 \left[ p_{N}^{\mu} p_1^{\nu} +  p_{N}^{\nu} p_1^{\mu} - g^{\mu \nu} p_N \cdot p_1 + i \eta \; \varepsilon^{\mu \nu \delta \eta} (p_N)_{\delta} (p_1)_{\eta} \right] \ .
\label{Lmunu}
\eea    
We use for $\gamma_5$ and $\varepsilon^{\mu \nu \delta \eta}$ the conventions of Ref.~\cite{IZ}.
It turns out that the $\eta$-dependence of the leptonic factor $L^{\mu \nu}$ and of the hadronic factor $H_{\mu} H_{\nu}^{*}$ cancel in the product (\ref{Tsq1}). If we define $\theta_1$ as\footnote{We recall: ${\vec p}_1$ is the 3-momentum in the $W^*=\ell_1 N$ rest frame $\Sigma$, and ${\vec q}'$ is the 3-momentum of $W^*$ in the $B$ rest sistem $\Sigma'$.} the angle between ${\vec p}_1$ and ${\hat z} = {\hat q}'$, this then leads after some algebra and summation over the polarization vectors of $D^{*}$ to the following differential decay rate:
 \bea
  \frac{d \Gamma (M \to V \ell_1 N)}{dq^2 d \Omega_{{\hat q}'} d \Omega_{{\hat p}_1}} & = &
  \frac{1}{8^4 \pi^5} \frac{|U_{\ell_1 N}|^2 |V_{qQ}|^2 G_F^2}{M_M^2}
  \blam^{1/2} 2 |{\vec {q'}}|  q^2 {\bigg \{}
\left[2 \left(1 - \frac{(M_N^2+M_1^2)}{q^2} \right) -
  \blam \sin^2 \theta_1  \right]
\left( ({\bar H_{+1}})^2 + ({\bar H_{-1}})^2 \right)
\nonumber\\
&&
- 2 \blam^{1/2} \cos \theta_1
\left( ({\bar H_{+1}})^2 - ({\bar H_{-1}})^2 \right)
+ 2 \left[ \left(1 - \frac{(M_N^2+M_1^2)}{q^2} \right) - \blam \cos^2 \theta_1 \right]
({\bar H^3})^2
\nonumber\\
&& +
4 \left( \frac{M_N^2-M_1^2}{q^2} \right) \blam^{1/2} \cos \theta_1 {\bar H^0}{\bar H^3}
+ 2 \left[
  - \left(\frac{M_N^2-M_1^2}{q^2} \right)^2 + \frac{(M_N^2+M_1^2)}{q^2}
  \right]
({\bar H^0})^2 {\bigg \}} \ .
    \label{dGdq2domdom2}
\eea
Here, $M_1$ is the mass of $\ell_1$, $V_{q Q}$ is the relevant CKM matrix element ($V_{q Q}=V _{cb}$ for $M=B, B_s, B_c$; $V_{q Q}=V_{c s}$ for $M=D_s$), and the following notations were used:
\bes
\label{notBDstellN}
\bea
|{\vec {q'}}| &=&
\frac{1}{2} M_M \lambda^{1/2}
\left( 1, \frac{ M_V^2}{M_M^2}, \frac{q^2}{M_M^2} \right),
\label{magq}
\\
\blam &\equiv&
\lambda
\left( 1, \frac{M_1^2}{q^2}, \frac{M_N^2}{q^2} \right) \ .
\label{blam}
\eea
\ees
and\footnote{
The use of the standard notation for $\lambda$ is made: 
\begin{displaymath}
\lambda^{1/2}(x,y,z) = \left[ x^2 + y^2 + z^2 - 2 x y - 2 y z - 2 z x \right]^{1/2}.
\end{displaymath}  
}
${\bar H}_{\pm 1}$, ${\bar H^0}$ and ${\bar H^3}$ denote the following expressions containing the mentioned form factors $V$ and $A_j$ ($j=0,1,2,3$):
\bes
 \label{bHs}
 \bea
     {\bar H_{\pm 1}} &=& (M_M+M_V) A_1(q^2)   \mp
     V(q^2) \frac{|{\vec {q'}}| 2 M_M}{(M_M+M_V)} \ ,
 \label{bHpm}
 \\
   {\bar H^3} & = & \frac{M_M^2}{2 M_V \sqrt{q^2}}
   \left[
          (M_M+M_V) A_1(q^2) \left(1 - \frac{(q^2+M_V^2)}{M_M^2} \right)
     - 4 A_2(q^2) \frac{|{\vec {q'}}|^2}{(M_M+M_V)}
     \right] \ ,
 \label{bH3}
 \\
   {\bar H^0} & = & \frac{M_M |{\vec {q'}}|}{M_V \sqrt{q^2}}
   \left[
     (M_M+M_V) A_1(q^2)  - (M_M- M_V) A_2(q^2) +
     2 M_V \left( A_0(q^2) - A_3(q^2) \right)
     \right] \ .
 \label{bH0}
\eea
\ees
In Appendix C of Ref.~\cite{CK1} we already obtained the differential decay rate (\ref{dGdq2domdom2}) [Eq.~(C19) there]; however, apart from some not relevant typos there, in the parametrization (\ref{FFBDst}) of the hadronic matrix elements we used the opposite sign in the first term there (proportional to $V$), which is inconvenient because it then corresponds to a negative form factor $V(q^2)$. As a consequence, the results of Appendix C of Ref.~\cite{CK1} should be reinterpreted with the substitution $V \mapsto -V$ in the formulas there. Therefore, the term proportional to $V$ in the differential decay rate (\ref{dGdq2domdom2}) here [i.e., the term with $({\bar H_{+1}})^2 - ({\bar H_{-1}})^2$] has now the sign opposite to that in Ref.~\cite{CK1}. Further, in Ref.~\cite{SLCK} we multiplied this term by $\eta$ ($=\pm 1$), the structure suggested in the literature (cf.,e.g., \cite{GiSi,RiBu}) which is obtained by keeping in the squared amplitude the $\eta$-dependence of the leptonic part $L^{\mu \nu}$ and regarding the hadronic part $H_{\mu} H^*_{\nu}$ as $\eta$-independent. We recall the the $\eta$-dependence of the hadronic elements, Eq.~(\ref{FFBDst}), is obtained by application of the charge conjugation transformation to the hadronic matrix elements. Furthermore, the measurements of the differential decay rates $B^+ \to {\bar D}^{*0} \ell_1^+ \nu_{\ell_1}$ \cite{Belle09} and ${\bar B}^0 \to D^{*+} \ell_1^- {\bar \nu}_{\ell_1}$ \cite{Belle18} (for which $M_N \approx 0 \approx M_1$) show that the differential rate is an increasing function of $\cos \theta_1$, i.e., that the sign in front of the term $\sim \cos \theta_1$ in Eq.~(\ref{dGdq2domdom2}) is negative\footnote{We recall that $({\bar H_{+1}})^2 - ({\bar H_{-1}})^2 \propto - A_1 V$ according to Eq.~(\ref{bHpm}).} in both types of decays, i.e., independent of $\eta$.

The use of the expression (\ref{dGdq2domdom2}) in the integration (\ref{Breff}) is in principle sensitive to the discussed term $\sim \cos \theta_1 (({\bar H_{+1}})^2 - ({\bar H_{-1}})^2)$, because of $\theta_1$-dependence of the decay probability $P_N$ there in the integrand. However, this term appears to affect the obtained upper bounds for $|U_{\ell N}|^2$ here insignificantly, by less than one per cent.

Furthermore, if we directly integrate the differential decay rate (\ref{dGdq2domdom2}) (as was the case in Ref.~\cite{CK1}, considering $P_N$ a constant), the discussed term $\sim \cos \theta_1$ gives exactly zero. Namely, in such a case,  the integration $d \Omega_{{\hat p}_1} = 2 \pi d \cos \theta_{\ell}$ can be performed explicitly, and the subsequent over $d \Omega_{{\hat q}'}$ gives then factor $4 \pi$, leading to the expression Eq.~(C20b) of Ref.~\cite{CK1}. The explicit expression for the total decay width $\Gamma(B \to D^{*} \ell_1 N)$ is then
\bea
\lefteqn{
{\Gamma}(M \to V \ell_1 N) =
\frac{|U_{\ell_1 N}|^2}{64 \pi^3} \frac{G_F^2 |V_{qQ}|^2}{M_M^2}
\int_{(M_N+M_1)^2}^{(M_M-M_V)^2}  d q^2 \;
\blam^{1/2} |{\vec q}| q^2 {\Bigg \{}
 \left( 1 - \frac{(M_N^2+M_1^2)}{q^2} - \frac{1}{3} \blam \right)
 {\bigg [} 2 (M_M+M_{\rm D})^2 A_1(q^2)^2
 }
 \nonumber\\
 &&
 + \frac{8 M_M^2 |{\vec q}|^2}{(M_M+M_V)^2} V(q^2)^2 +  \frac{M_M^4}{4 M_V^2 q^2} \left( (M_M+M_V)
   \left( 1 - \frac{(q^2+M_V^2)}{M_M^2} \right) A_1(q^2) -  \frac{4 |{\vec q}|^2}{(M_M+M_V)} A_2(q^2) \right)^2 {\bigg ]}
 \nonumber\\
 &&
 + \left[ - \left(\frac{M_N^2-M_1^2}{q^2} \right)^2 + \frac{(M_N^2+M_1^2)}{q^2} \right] \frac{4 M_M^2 |{\vec q}|^2}{q^2} A_0(q^2)^2 {\Bigg \}}.
 \label{bGBDstNl}
 \eea
The explicit expression for this total decay width in Ref.~\cite{CK1} [Eq.~(19) there] was written for the case of an approximation of the form factor $A_0$ as a combination of $A_1$ and $A_2$ [Eq.~(17) there], but is written here, for completeness, in the form independent of this approximation.

\section*{Acknowledgments}
\noindent
The work of G.C. was supported in part by FONDECYT (Chile) Grant No.~1180344; the work of C.S.K. was supported in part by the National Research Foundation of Korea (NRF) grant funded by the Korean government (MSIP) (NRF-2018R1A4A1025334). We thank Sheldon L.~Stone for providing us with valuable information on the LHCb upgrade experiment.


\begin{thebibliography}{99}

\bibitem{0nubb}
  G.~Racah,
  ``On the symmetry of particle and antiparticle,''
  Nuovo Cimento  {\bf 14}, 322 (1937);
  W.~H.~Furry,
  ``On transition probabilities in double beta-disintegration,''
  Phys.\ Rev.\  {\bf 56}, 1184 (1939);
H.~Primakoff and S.~P.~Rosen, ``Double beta decay,''
Rep. Prog. Phys. {\bf 22}, 121 (1959);
  ``Nuclear double-beta decay and a new limit on lepton nonconservation,''
  Phys.\ Rev.\  {\bf 184}, 1925 (1969);
  ``Baryon number and lepton number conservation laws,''
  Annu.\ Rev.\ Nucl.\ Part.\ Sci.\  {\bf 31}, 145 (1981);
  J.~Schechter and J.~W.~F.~Valle,
  ``Neutrinoless double beta decay in $SU(2) x U(1)$ theories,''
  Phys.\ Rev.\ D {\bf 25}, 2951 (1982);
 M.~Doi, T.~Kotani and E.~Takasugi,
  ``Double beta decay and Majorana neutrino,''
  Prog.\ Theor.\ Phys.\ Suppl.\  {\bf 83}, 1 (1985);
     S.~R.~Elliott and J.~Engel,
  ``Double beta decay,''
  J.\ Phys.\ G  {\bf 30}, R183 (2004)
  [hep-ph/0405078];
  V.~A.~Rodin, A.~Faessler, F.~\v{S}imkovic and P.~Vogel,
  ``Assessment of uncertainties in QRPA $0\nu\beta\beta$-decay nuclear matrix elements,''
  Nucl.\ Phys.\ A {\bf 766}, 107 (2006);
  Erratum, Nucl.\ Phys.\ A {\bf 793}, 213(E) (2007)
  [arXiv:0706.4304 [nucl-th]].

\bibitem{scatt1}
  W.~-Y.~Keung and G.~Senjanovi\'c,
  ``Majorana neutrinos and the production of the right-handed charged gauge boson,''
  Phys.\ Rev.\ Lett.\  {\bf 50}, 1427 (1983);
  V.~Tello, M.~Nemev\v{s}ek, F.~Nesti, G.~Senjanovi\'c and F.~Vissani,
  ``Left-right symmetry: from LHC to Neutrinoless double beta decay,''
  Phys.\ Rev.\ Lett.\  {\bf 106}, 151801 (2011)
 [arXiv:1011.3522 [hep-ph]];
  M.~Nemev\v{s}ek, F.~Nesti, G.~Senjanovi\'c and V.~Tello,
  ``Neutrinoless double beta decay: low left-right symmetry scale?,''
  arXiv:1112.3061 [hep-ph];
  S.~Kovalenko, Z.~Lu and I.~Schmidt,
  ``Lepton number violating processes mediated by Majorana neutrinos at hadron colliders,''
  Phys.\ Rev.\ D {\bf 80}, 073014 (2009)
  [arXiv:0907.2533 [hep-ph]];
  J.~C.~Helo, M.~Hirsch and S.~Kovalenko,
  ``Heavy neutrino searches at the LHC with displaced vertices,''
  Phys.\ Rev.\ D {\bf 89}, 073005 (2014);
  Erratum, Phys.\ Rev.\ D {\bf 93}, no. 9, 099902(E) (2016)
  [arXiv:1312.2900 [hep-ph]];
  G.~Senjanovi\'c,
  ``Neutrino mass: From LHC to grand unification,''
 Riv.\ Nuovo Cim.\  {\bf 34}, 1 (2011);
  C.~Y.~Chen and P.~S.~Bhupal Dev,
  ``Multi-lepton collider signatures of heavy Dirac and Majorana neutrinos,''
  Phys.\ Rev.\ D {\bf 85}, 093018 (2012)
  [arXiv:1112.6419 [hep-ph]];
  C.~Y.~Chen, P.~S.~Bhupal Dev and R.~N.~Mohapatra,
  ``Probing Heavy-Light Neutrino Mixing in Left-Right Seesaw Models at the LHC,''
  Phys.\ Rev.\ D {\bf 88}, 033014 (2013)
  [arXiv:1306.2342 [hep-ph]];
  P.~S.~Bhupal Dev, A.~Pilaftsis and U.~k.~Yang,
  ``New Production Mechanism for Heavy Neutrinos at the LHC,''
  Phys.\ Rev.\ Lett.\  {\bf 112}, 081801 (2014)
  [arXiv:1308.2209 [hep-ph]];
  A.~Das and N.~Okada,
  ``Inverse seesaw neutrino signatures at the LHC and ILC,''
  Phys.\ Rev.\ D {\bf 88}, 113001 (2013)
  [arXiv:1207.3734 [hep-ph]];
  A.~Das, P.~S.~Bhupal Dev and N.~Okada,
  ``Direct bounds on electroweak scale pseudo-Dirac neutrinos from $\sqrt s=8$ TeV LHC data,''
  Phys.\ Lett.\ B {\bf 735}, 364 (2014)
  [arXiv:1405.0177 [hep-ph]];
  D.~Alva, T.~Han and R.~Ruiz,
  ``Heavy Majorana neutrinos from $W\gamma$ fusion at hadron colliders,''
  JHEP {\bf 1502}, 072 (2015)
  [arXiv:1411.7305 [hep-ph]];
  A.~Das and N.~Okada,
  ``Improved bounds on the heavy neutrino productions at the LHC,''
  Phys.\ Rev.\ D {\bf 93}, no. 3, 033003 (2016)
  [arXiv:1510.04790 [hep-ph]];
  ``Bounds on heavy Majorana neutrinos in type-I seesaw and implications for collider searches,''
  Phys.\ Lett.\ B {\bf 774}, 32 (2017)
  [arXiv:1702.04668 [hep-ph]];
  C.~Degrande, O.~Mattelaer, R.~Ruiz and J.~Turner,
  ``Fully-automated precision predictions for heavy neutrino production mechanisms at hadron colliders,''
  Phys.\ Rev.\ D {\bf 94}, no. 5, 053002 (2016)
  [arXiv:1602.06957 [hep-ph]];
  A.~Das, P.~Konar and S.~Majhi,
  ``Production of Heavy neutrino in next-to-leading order QCD at the LHC and beyond,''
  JHEP {\bf 1606}, 019 (2016)
  [arXiv:1604.00608 [hep-ph]];
  A.~Das,
  ``Pair production of heavy neutrinos in next-to-leading order QCD at the hadron colliders in the inverse seesaw framework,''
  arXiv:1701.04946 [hep-ph];
  L.~Duarte, J.~Peressutti and O.~A.~Sampayo,
  ``Not-that-heavy Majorana neutrino signals at the LHC,''
  J.\ Phys.\ G {\bf 45}, no. 2, 025001 (2018)
  [arXiv:1610.03894 [hep-ph]].

  \bibitem{scatt2}
  W.~Buchm\"uller and C.~Greub,
  ``Heavy Majorana neutrinos in electron - positron and electron - proton collisions,''
  Nucl.\ Phys.\ B {\bf 363}, 345 (1991);
  M.~Kohda, H.~Sugiyama and K.~Tsumura,
  ``Lepton number violation at the LHC with leptoquark and diquark,''
  Phys.\ Lett.\ B {\bf 718}, 1436 (2013)
  [arXiv:1210.5622 [hep-ph]].




  
  \bibitem{mesdec1}
 L.~S.~Littenberg and R.~E.~Shrock,
  ``Upper bounds on lepton number violating meson decays,''
  Phys.\ Rev.\ Lett.\  {\bf 68}, 443 (1992);
  ``Implications of improved upper bounds on $|\Delta L| = 2$ processes,''
  Phys.\ Lett.\ B {\bf 491}, 285 (2000)
  [hep-ph/0005285];
 C.~Dib, V.~Gribanov, S.~Kovalenko and I.~Schmidt,
  ``K meson neutrinoless double muon decay as a probe of neutrino masses and mixings,''
  Phys.\ Lett.\ B {\bf 493}, 82 (2000)
  [hep-ph/0006277];
  A.~Ali, A.~V.~Borisov and N.~B.~Zamorin,
  ``Majorana neutrinos and same sign dilepton production at LHC and in rare meson decays,''
  Eur.\ Phys.\ J.\ C {\bf 21}, 123 (2001)
  [hep-ph/0104123];
  M.~A.~Ivanov and S.~G.~Kovalenko,
  ``Hadronic structure aspects of $K^{+} \to \pi^- + l^+_1 + l^+_2$ decays,''
  Phys.\ Rev.\ D {\bf 71}, 053004 (2005)
  [hep-ph/0412198];
  A.~de Gouvea and J.~Jenkins,
  ``Survey of lepton number violation via effective operators,''
  Phys.\ Rev.\ D {\bf 77}, 013008 (2008)
 [arXiv:0708.1344 [hep-ph]];
  N.~Quintero, G.~L\'opez Castro and D.~Delepine,
  ``Lepton number violation in top quark and neutral B meson decays,''
  Phys.\ Rev.\ D {\bf 84}, 096011 (2011);
  Erratum, Phys.\ Rev.\ D {\bf 86}, 079905(E) (2012)
  [arXiv:1108.6009 [hep-ph]];
  G.~L.~Castro and N.~Quintero,
  ``Bounding resonant Majorana neutrinos from four-body B and D decays,''
  Phys.\ Rev.\ D {\bf 87}, 077901 (2013)
  [arXiv:1302.1504 [hep-ph]];
  A.~Abada, A.~M.~Teixeira, A.~Vicente and C.~Weiland,
  ``Sterile neutrinos in leptonic and semileptonic decays,''
  JHEP {\bf 1402}, 091 (2014)
  [arXiv:1311.2830 [hep-ph]];
  Y.~Wang, S.~S.~Bao, Z.~H.~Li, N.~Zhu and Z.~G.~Si,
  ``Study Majorana neutrino contribution to B-meson semi-leptonic rare decays,''
  Phys.\ Lett.\ B {\bf 736}, 428 (2014)
  [arXiv:1407.2468 [hep-ph]].

\bibitem{Atre}  
  A.~Atre, T.~Han, S.~Pascoli and B.~Zhang,
  ``The search for heavy Majorana neutrinos,''
  JHEP {\bf 0905}, 030 (2009)
  [arXiv:0901.3589 [hep-ph]].

\bibitem{SKetal} 
  J.~C.~Helo, S.~Kovalenko and I.~Schmidt,
  ``Sterile neutrinos in lepton number and lepton flavor violating decays,''
  Nucl.\ Phys.\ B {\bf 853}, 80 (2011)
  [arXiv:1005.1607 [hep-ph]].

\bibitem{mesdecOur}
  G.~Cveti\v{c}, C.~Dib, S.~K.~Kang and C.~S.~Kim,
  ``Probing Majorana neutrinos in rare $K$ and $D, ~D_s, B, B_c$ meson decays,''
  Phys.\ Rev.\ D {\bf 82}, 053010 (2010)
  [arXiv:1005.4282 [hep-ph]];
  G.~Cveti\v{c}, C.~Dib and C.~S.~Kim,
  ``Probing Majorana neutrinos in rare $\pi^+ \to e^+ e^+ \mu^- \nu$ decays,''
  JHEP {\bf 1206}, 149 (2012)
  [arXiv:1203.0573 [hep-ph]].

\bibitem{symm}
  G.~Cveti\v{c}, C.~Dib, C.~S.~Kim and J.~Zamora-Sa\'a,
  ``Probing the Majorana neutrinos and their CP violation in decays of charged scalar mesons $\pi, K, D, D_s, B, B_c$,''
  Symmetry {\bf 7}, 726 (2015)
  [arXiv:1503.01358 [hep-ph]].

\bibitem{mesdecnew}  
  T.~Asaka and H.~Ishida,
  ``Lepton number violation by heavy Majorana neutrino in $B$ decays,''
  Phys.\ Lett.\ B {\bf 763}, 393 (2016)
  [arXiv:1609.06113 [hep-ph]];
  S.~Mandal and N.~Sinha,
  ``Favoured $B_c$ decay modes to search for a Majorana neutrino,''
  Phys.\ Rev.\ D {\bf 94}, no. 3, 033001 (2016)
  [arXiv:1602.09112 [hep-ph]];
  J.~Mej\'{\i}a-Guisao, D.~Milan\'es, N.~Quintero and J.~D.~Ruiz-\'Alvarez,
  ``Exploring GeV-scale Majorana neutrinos in lepton-number-violating $\Lambda_b^0$ baryon decays,''
  Phys.\ Rev.\ D {\bf 96}, no. 1, 015039 (2017)
  [arXiv:1705.10606 [hep-ph]];
  D.~Milan\'es and N.~Quintero,
  ``Search for lepton-number-violating signals in the charm sector,''
  Phys.\ Rev.\ D {\bf 98}, no. 9, 096004 (2018)
  [arXiv:1808.06017 [hep-ph]];
  H.~Yuan, T.~Wang, Y.~Jiang, Q.~Li and G.~L.~Wang,
  ``Four-body decays of $B$ meson with lepton number violation,''
  J.\ Phys.\ G {\bf 45}, no. 6, 065002 (2018)
  [arXiv:1710.03886 [hep-ph]];
  C.~X.~Yue and J.~P.~Chu,
  ``Sterile neutrino and leptonic decays of the pseudoscalar mesons,''
  Phys.\ Rev.\ D {\bf 98}, no. 5, 055012 (2018)
  [arXiv:1808.09139 [hep-ph]];
  S.~Hu, S.~M.~Y.~Wong and F.~Xu,
  ``Probing sterile neutrino via lepton flavor violating decays of mesons,''
  arXiv:1904.00568 [hep-ph];
  L.~Duarte, J.~Peressutti, I.~Romero and O.~A.~Sampayo,
  ``Majorana neutrinos with effective interactions in B decays,''
  arXiv:1904.07175 [hep-ph].

\bibitem{mesdecQuint}
  D.~Milan\'es, N.~Quintero and C.~E.~Vera,
  ``Sensitivity to Majorana neutrinos in $\Delta L=2$ decays of $B_c$ meson at LHCb,''
  Phys.\ Rev.\ D {\bf 93}, no. 9, 094026 (2016)
  [arXiv:1604.03177 [hep-ph]];
  J.~Mej\'{\i}a-Guisao, D.~Milan\'es, N.~Quintero and J.~D.~Ruiz-\'Alvarez,
  ``Lepton number violation in $B_s$ meson decays induced by an on-shell Majorana neutrino,''
  Phys.\ Rev.\ D {\bf 97}, no. 7, 075018 (2018)
  [arXiv:1708.01516 [hep-ph]].

\bibitem{mesdecShrock}
  D.~A.~Bryman and R.~Shrock,
  ``Improved constraints on sterile neutrinos in the MeV to GeV mass range,''
  arXiv:1904.06787 [hep-ph].

\bibitem{mestaudec}
  K.~Bondarenko, A.~Boyarsky, D.~Gorbunov and O.~Ruchayskiy,
  ``Phenomenology of GeV-scale heavy neutral leptons,''
  JHEP {\bf 1811}, 032 (2018)
  [arXiv:1805.08567 [hep-ph]].
  
\bibitem{taudec}
  C.~S.~Kim, G.~L\'opez Castro and D.~Sahoo,
  ``Discovering intermediate mass sterile neutrinos through $\tau^- \to \pi^- \mu^- e^+ \nu$ (or $\bar{\nu}$) decay,''
  Phys.\ Rev.\ D {\bf 96}, no. 7, 075016 (2017)
  [arXiv:1708.00802 [hep-ph]];
  A.~Abada, V.~De Romeri, M.~Lucente, A.~M.~Teixeira and T.~Toma,
  ``Effective Majorana mass matrix from tau and pseudoscalar meson lepton number violating decays,''
  JHEP {\bf 1802}, 169 (2018)
  [arXiv:1712.03984 [hep-ph]].

\bibitem{CPtau}
  J.~Zamora-Sa\'a,
  ``Resonant $CP$ violation in rare $\tau^{\pm}$ decays,''
  JHEP {\bf 1705}, 110 (2017)
  [arXiv:1612.07656 [hep-ph]].

\bibitem{Wdec}
  C.~O.~Dib, C.~S.~Kim, N.~A.~Neill and X.~B.~Yuan,
  ``Search for sterile neutrinos decaying into pions at the LHC,''
  Phys.\ Rev.\ D {\bf 97}, no. 3, 035022 (2018)
  [arXiv:1801.03624 [hep-ph]];
  C.~O.~Dib, C.~S.~Kim and S.~Tapia Araya,
  ``Search for light sterile neutrinos from $W^\pm$ decays at the LHC,''
  arXiv:1903.04905 [hep-ph];
  M.~Drewes and J.~Hajer,
  ``Heavy neutrinos in displaced vertex searches at the LHC and HL-LHC,''
  arXiv:1903.06100 [hep-ph];
  K.~Bondarenko, A.~Boyarsky, M.~Ovchynnikov, O.~Ruchayskiy and L.~Shchutska,
  ``Probing new physics with displaced vertices: muon tracker at CMS,''
  arXiv:1903.11918 [hep-ph];
  J.~Liu, Z.~Liu, L.~T.~Wang and X.~P.~Wang,
  ``Seeking for sterile neutrinos with displaced leptons at the LHC,''
  arXiv:1904.01020 [hep-ph].
  
\bibitem{Zdec}
  A.~Das, S.~Jana, S.~Mandal and S.~Nandi,
  ``Probing right handed neutrinos at the LHeC and lepton colliders using fat jet signatures,''
  Phys.\ Rev.\ D {\bf 99}, no. 5, 055030 (2019)
  [arXiv:1811.04291 [hep-ph]];
  J.~N.~Ding, Q.~Qin and F.~S.~Yu,
  ``Heavy neutrino searches at future $Z$-factories,''
  arXiv:1903.02570 [hep-ph].

\bibitem{BhaGao}
  B.~Bhattacharya, C.~M.~Grant and A.~A.~Petrov,
  ``Invisible widths of heavy mesons,''
  Phys.\ Rev.\ D {\bf 99}, no. 9, 093010 (2019)
  [arXiv:1809.04606 [hep-ph]];
  D.~N.~Gao,
  ``Note on invisible decays of light mesons,''
  Phys.\ Rev.\ D {\bf 98}, no. 11, 113006 (2018)
  [arXiv:1811.10152 [hep-ph]].
  
\bibitem{neuosc}
  B.~Pontecorvo,
  ``Inverse beta processes and nonconservation of lepton charge,''
  Zh.\ Eksp.\ Teor.\ Fiz.\  {\bf 34}, 247 (1957) [Sov.\ Phys.\ JETP {\bf 7}, 172 (1958)];
  ``Neutrino experiments and the problem of conservation of leptonic charge,''
 Zh.\ Eksp.\ Teor.\ Fiz.\  {\bf 53}, 1717 (1967)  [Sov.\ Phys.\ JETP {\bf 26}, 984 (1968)].

  
\bibitem{oscobs}
  Y.~Fukuda {\it et al.}  [Super-Kamiokande Collaboration],
  ``Evidence for oscillation of atmospheric neutrinos,''
  Phys.\ Rev.\ Lett.\  {\bf 81}, 1562 (1998)
  [hep-ex/9807003];
  Q.~R.~Ahmad {\it et al.}  [SNO Collaboration],
  ``Direct evidence for neutrino flavor transformation from neutral current interactions in the Sudbury Neutrino Observatory,''
  Phys.\ Rev.\ Lett.\  {\bf 89}, 011301 (2002)
  [nucl-ex/0204008];
  K.~Eguchi {\it et al.}  [KamLAND Collaboration],
  ``First results from KamLAND: Evidence for reactor anti-neutrino disappearance,''
  Phys.\ Rev.\ Lett.\  {\bf 90}, 021802 (2003)
  [hep-ex/0212021].
  
\bibitem{oscheavy}
  D.~Boyanovsky,
  ``Nearly degenerate heavy sterile neutrinos in cascade decay: mixing and oscillations,''
  Phys.\ Rev.\ D {\bf 90}, 105024 (2014)
  [arXiv:1409.4265 [hep-ph]];
  G.~Cveti\v{c}, C.~S.~Kim, R.~K\"ogerler and J.~Zamora-Sa\'a,
  ``Oscillation of heavy sterile neutrino in decay of $B \to \mu e \pi$,''
  Phys.\ Rev.\ D {\bf 92}, 013015 (2015)
  [arXiv:1505.04749 [hep-ph]];
  G.~Cveti\v{c}, A.~Das and J.~Zamora-Sa\'a,
  ``Probing heavy neutrino oscillations in rare W boson decays,''
  J. Phys. G (2019)
  https://doi.org/10.1088/1361-6471/ab1212
  [arXiv:1805.00070 [hep-ph]];
  S.~Antusch, E.~Cazzato and O.~Fischer,
  ``Resolvable heavy neutrino antineutrino oscillations at colliders,''
  Mod.\ Phys.\ Lett.\ A {\bf 34}, no. 07n08, 1950061 (2019)
  [arXiv:1709.03797 [hep-ph]].
  
\bibitem{CPscatt}
  A.~Pilaftsis,
  ``CP violation and baryogenesis due to heavy Majorana neutrinos,''
  Phys.\ Rev.\ D {\bf 56}, 5431 (1997)
  [hep-ph/9707235];
  S.~Bray, J.~S.~Lee and A.~Pilaftsis,
  ``Resonant CP violation due to heavy neutrinos at the LHC,''
  Nucl.\ Phys.\ B {\bf 786}, 95 (2007)
  [hep-ph/0702294 [HEP-PH]].

  
\bibitem{CKZCP1}
  G.~Cveti\v{c}, C.~S.~Kim and J.~Zamora-Sa\'a,
  ``CP violations in $\pi^{\pm}$ meson decay,''
  J.\ Phys.\ G {\bf 41}, 075004 (2014)
  [arXiv:1311.7554 [hep-ph]].
  
\bibitem{CKZCP2}
  G.~Cveti\v{c}, C.~S.~Kim and J.~Zamora-Sa\'a,
 ``CP violation in lepton number violating semihadronic decays of $K,D,D_s,B,B_c$,''
  Phys.\ Rev.\ D {\bf 89}, no. 9, 093012 (2014)
  [arXiv:1403.2555 [hep-ph]].

\bibitem{DCK}
  C.~O.~Dib, M.~Campos and C.~S.~Kim,
  ``CP violation with Majorana neutrinos in $K$ meson decays,''
  JHEP {\bf 1502}, 108 (2015)
  [arXiv:1403.8009 [hep-ph]].
  
\bibitem{lsseesaw}
 L.~Canetti, M.~Drewes and B.~Garbrecht,
``Probing leptogenesis with GeV-scale sterile neutrinos at LHCb and Belle II,''
Phys.\ Rev.\ D {\bf 90}, 125005 (2014)
 doi:10.1103/PhysRevD.90.125005
  [arXiv:1404.7114 [hep-ph]];
  M.~Drewes and B.~Garbrecht,
``Experimental and cosmological constraints on heavy neutrinos,''
arXiv:1502.00477 [hep-ph];
  G.~Moreno and J.~Zamora-Sa\'a,
  ``Rare meson decays with three pairs of quasi-degenerate heavy neutrinos,''
  Phys.\ Rev.\ D {\bf 94}, no. 9, 093005 (2016)
  doi:10.1103/PhysRevD.94.093005
  [arXiv:1606.08820 [hep-ph]].
  
\bibitem{nuMSM1}
  T.~Asaka, S.~Blanchet and M.~Shaposhnikov,
  ``The $\nu$MSM, dark matter and neutrino masses,''
  Phys.\ Lett.\ B {\bf 631}, 151 (2005)
  [hep-ph/0503065];
  T.~Asaka and M.~Shaposhnikov,
  ``The $\nu$MSM, dark matter and baryon asymmetry of the universe,''
  Phys.\ Lett.\ B {\bf 620}, 17 (2005)
  [hep-ph/0505013].
  
\bibitem{nuMSM2}
 D.~Gorbunov and M.~Shaposhnikov,
  ``How to find neutral leptons of the $\nu$MSM?,''
 JHEP {\bf 0710}, 015 (2007);
 Erratum, JHEP {\bf 1311}, 101(E) (2013)
  [arXiv:0705.1729 [hep-ph]];
  A.~Boyarsky, O.~Ruchayskiy and M.~Shaposhnikov,
  ``The role of sterile neutrinos in cosmology and astrophysics,''
  Annu.\ Rev.\ Nucl.\ Part.\ Sci.\  {\bf 59}, 191 (2009)
  [arXiv:0901.0011 [hep-ph]];
  L.~Canetti, M.~Drewes and M.~Shaposhnikov,
  ``Sterile neutrinos as the origin of dark and baryonic matter,''
  Phys.\ Rev.\ Lett.\  {\bf 110}, 061801 (2013)
  [arXiv:1204.3902 [hep-ph]];
  L.~Canetti, M.~Drewes, T.~Frossard and M.~Shaposhnikov,
  ``Dark matter, baryogenesis and neutrino oscillations from right handed neutrinos,''
  Phys.\ Rev.\ D {\bf 87}, 093006 (2013)
  [arXiv:1208.4607 [hep-ph]].
  
\bibitem{seesawLNV}
  Y.~Cai, T.~Han, T.~Li and R.~Ruiz,
  ``Lepton number violation: seesaw models and their collider tests,''
  Front.\ in Phys.\  {\bf 6}, 40 (2018)
  [arXiv:1711.02180 [hep-ph]].

\bibitem{seesaw}
  P.~Minkowski,
  ``$\mu \to e \gamma$ at a rate of one out of $10^9$ muon decays?,''
  Phys.\ Lett.\ B {\bf 67}, 421 (1977)
 doi:10.1016/0370-2693(77)90435-X;
M.~Gell-Mann, P.~Ramond and R.~Slansky, in Sanibel Conference,
``The family group in Grand Unified Theories,'' Febr.~1979, Report No.~CALT-68-709, reprinted in hep-ph/9809459;
``Complex spinors and unified theories,'' Print 80-0576,
published in: D.~Freedman et al. (Eds.), {\it Supergravity}, North-Holland, Amsterdam, 1979;
  T.~Yanagida,
  ``Horizontal symmetry and masses of neutrinos,''
  Conf.\ Proc.\ C {\bf 7902131}, 95 (1979);
S.~L.~Glashow, in: M.~Levy et al. (Eds.), {\it Quarks and Leptons}, Cargese,
Plenum, New York, 1980, p.~707;
  R.~N.~Mohapatra and G.~Senjanovi\'c,
  ``Neutrino mass and spontaneous parity violation,''
  Phys.\ Rev.\ Lett.\  {\bf 44}, 912 (1980).
  doi:10.1103/PhysRevLett.44.912

\bibitem{seesaw1TeV}
  D.~Wyler and L.~Wolfenstein,
  ``Massless neutrinos in left-right symmetric models,''
  Nucl.\ Phys.\ B {\bf 218}, 205 (1983);
  E.~Witten,
  ``Symmetry breaking patterns in superstring models,''
  Nucl.\ Phys.\ B {\bf 258}, 75 (1985);
  R.~N.~Mohapatra and J.~W.~F.~Valle,
  ``Neutrino mass and baryon number nonconservation in superstring models,''
  Phys.\ Rev.\ D {\bf 34}, 1642 (1986);
  M.~Malinsky, J.~C.~Romao and J.~W.~F.~Valle,
  ``Novel supersymmetric $SO(10)$ seesaw mechanism,''
  Phys.\ Rev.\ Lett.\  {\bf 95}, 161801 (2005)
  [hep-ph/0506296];
  P.~S.~Bhupal Dev and R.~N.~Mohapatra,
  ``TeV scale inverse seesaw in $SO(10)$ and leptonic non-unitarity effects,''
  Phys.\ Rev.\ D {\bf 81}, 013001 (2010)
  [arXiv:0910.3924 [hep-ph]];
  P.~S.~Bhupal Dev and A.~Pilaftsis,
  ``Minimal radiative neutrino mass mechanism for inverse seesaw models,''
  Phys.\ Rev.\ D {\bf 86}, 113001 (2012)
  [arXiv:1209.4051 [hep-ph]];
  C.~H.~Lee, P.~S.~Bhupal Dev and R.~N.~Mohapatra,
  ``Natural TeV-scale left-right seesaw mechanism for neutrinos and experimental tests,''
  Phys.\ Rev.\ D {\bf 88}, 093010 (2013)
  [arXiv:1309.0774 [hep-ph]].

\bibitem{seesaw1GeV}  
  T.~Appelquist and R.~Shrock,
  ``Neutrino masses in theories with dynamical electroweak symmetry breaking,''
  Phys.\ Lett.\ B {\bf 548}, 204 (2002)
  [hep-ph/0204141];
  ``Dynamical symmetry breaking of extended gauge symmetries,''
  Phys.\ Rev.\ Lett.\  {\bf 90}, 201801 (2003)
  [hep-ph/0301108];
  ``Fermion masses and mixing in extended technicolor models,''
  Phys.\ Rev.\ D {\bf 69}, 015002 (2004)
  [hep-ph/0308061];
  F.~del Aguila, J.~A.~Aguilar-Saavedra, J.~de Blas and M.~Zralek,
  ``Looking for signals beyond the neutrino Standard Model,''
  Acta Phys.\ Polon.\ B {\bf 38}, 3339 (2007)
  [arXiv:0710.2923 [hep-ph]];
  X.~G.~He, S.~Oh, J.~Tandean and C.~C.~Wen,
  ``Large mixing of light and heavy neutrinos in seesaw models and the LHC,''
  Phys.\ Rev.\ D {\bf 80}, 073012 (2009)
  [arXiv:0907.1607 [hep-ph]];
  J.~Kersten and A.~Y.~Smirnov,
  ``Right-handed neutrinos at CERN LHC and the mechanism of neutrino mass generation,''
  Phys.\ Rev.\ D {\bf 76}, 073005 (2007)
  [arXiv:0705.3221 [hep-ph]];
  A.~Ibarra, E.~Molinaro and S.~T.~Petcov,
  ``TeV scale see-saw mechanisms of neutrino mass generation, the Majorana nature of the heavy singlet neutrinos and $0\nu\beta\beta$-decay,''
  JHEP {\bf 1009}, 108 (2010)
  [arXiv:1007.2378 [hep-ph]];
  M.~Nemev\v{s}ek, G.~Senjanovi\'c and Y.~Zhang,
  ``Warm dark matter in low scale left-right theory,''
  JCAP {\bf 1207}, 006 (2012)
  [arXiv:1205.0844 [hep-ph]].


  
\bibitem{SLCK}
  G.~Cveti\v{c} and C.~S.~Kim,
  ``Sensitivity limits on heavy-light mixing $|U_{\mu N}|^2$ from lepton number violating $B$ meson decays,''
  Phys.\ Rev.\ D {\bf 96}, no. 3, 035025 (2017)
  doi:10.1103/PhysRevD.96.035025
  [arXiv:1705.09403 [hep-ph]].
  
\bibitem{PDG2018}
  M.~Tanabashi {\it et al.} [Particle Data Group],
  Phys.\ Rev.\ D {\bf 98}, no. 3, 030001 (2018).

\bibitem{CK1}  
  G.~Cveti\v{c} and C.~S.~Kim,
  ``Rare decays of B mesons via on-shell sterile neutrinos,''
  Phys.\ Rev.\ D {\bf 94}, no. 5, 053001 (2016);
  Erratum, Phys.\ Rev.\ D {\bf 95}, no. 3, 039901(E) (2017)
  [arXiv:1606.04140 [hep-ph]].

  

  \bibitem{VELO}
  R.~Aaij {\it et al.} [LHCb Collaboration],
  ``LHCb detector performance,''
  Int.\ J.\ Mod.\ Phys.\ A {\bf 30}, no. 07, 1530022 (2015)
  [arXiv:1412.6352 [hep-ex]];
  ``LHCb VELO Upgrade Technical Design Report,'' CERN-LHCC-2013-021,  29 Nov.~2013, https://cds.cern.ch/record/1624070, cf.~Fig.~4 there.

 \bibitem{Sheldon}
 Sheldon L.~Stone, private communication.
 
\bibitem{CKWW}
  G.~Cveti\v{c}, C.~S.~Kim, G.~L.~Wang and W.~Namgung,
  ``Decay constants of heavy meson of 0- state in relativistic Salpeter method,''
  Phys.\ Lett.\ B {\bf 596}, 84 (2004)
  [hep-ph/0405112].
  
 
 \bibitem{CLN}
  I.~Caprini, L.~Lellouch and M.~Neubert,
  ``Dispersive bounds on the shape of ${\overline B} \to D^{(*)} \ell {\overline \nu}$ form factors,''
  Nucl.\ Phys.\ B {\bf 530}, 153 (1998)
  [hep-ph/9712417].

\bibitem{Belle15}
  R.~Glattauer {\it et al.} [Belle Collaboration],
  ``Measurement of the decay $B\to D\ell\nu_\ell$ in fully reconstructed events and determination of the Cabibbo-Kobayashi-Maskawa matrix element $|V_{cb}|$,''
  Phys.\ Rev.\ D {\bf 93}, no. 3, 032006 (2016)
  [arXiv:1510.03657 [hep-ex]].

\bibitem{Belle2}
  W.~Dungel {\it et al.} [Belle Collaboration],
  ``Measurement of the form factors of the decay $B^0 \to D^{*-} \ell^{+} \nu$ and determination of the CKM matrix element $|V_{cb}|$,''
  Phys.\ Rev.\ D {\bf 82}, 112007 (2010)
  [arXiv:1010.5620 [hep-ex]].
  
\bibitem{Fanetal}
  Y.~Y.~Fan, W.~F.~Wang and Z.~J.~Xiao,
  ``Study of $\bar{B}_s^0 \to (D_s^+,D_s^{*+}) l^-\bar{\nu}_l$ decays in the pQCD factorization approach,''
  Phys.\ Rev.\ D {\bf 89}, no. 1, 014030 (2014)
  [arXiv:1311.4965 [hep-ph]].

\bibitem{Wangetal}
  W.~Wang, Y.~L.~Shen and C.~D.~Lu,
  ``Covariant Light-Front Approach for $B_c$ transition form factors,''
  Phys.\ Rev.\ D {\bf 79}, 054012 (2009)
  [arXiv:0811.3748 [hep-ph]].
  
\bibitem{DM}
  G.~Duplan\v{c}i\'c and B.~Meli\'c,
  ``Form factors of $B, B_{s} \to \eta^{(')}$ and $D, D_{s} \to \eta^{(')}$ transitions from QCD light-cone sum rules,''
  JHEP {\bf 1511}, 138 (2015)
  [arXiv:1508.05287 [hep-ph]].

\bibitem{WSh}
  W.~Wang and Y.~L.~Shen,
  ``$D_ss \to K, K^{*}, \phi$ form factors in the Covariant Light-Front approach and exclusive $D_s$ decays,''
  Phys.\ Rev.\ D {\bf 78}, 054002 (2008).

  \bibitem{FC}
  G.~J.~Feldman and R.~D.~Cousins,
  ``A unified approach to the classical statistical analysis of small signals,''
  Phys.\ Rev.\ D {\bf 57}, 3873 (1998)
  [physics/9711021 [physics.data-an]].
  
\bibitem{PS191}
  G.~Bernardi {\it et al.},
  ``Further limits on heavy neutrino couplings,''
  Phys.\ Lett.\ B {\bf 203}, 332 (1988).
  
\bibitem{NuTeV}
  A.~Vaitaitis {\it et al.} [NuTeV and E815 Collaborations],
  ``Search for neutral heavy leptons in a high-energy neutrino beam,''
  Phys.\ Rev.\ Lett.\  {\bf 83}, 4943 (1999)
  [hep-ex/9908011].

\bibitem{NA3}
  J.~Badier {\it et al.} [NA3 Collaboration],
  ``Mass and lifetime limits on new longlived particles in 300 GeV  $\pi^-$ interactions,''
  Z.\ Phys.\ C {\bf 31}, 21 (1986).

\bibitem{BEBC}
  A.~M.~Cooper-Sarkar {\it et al.} [WA66 Collaboration],
  ``Search for heavy neutrino decays in the BEBC beam dump experiment,''
  Phys.\ Lett.\  {\bf 160B}, 207 (1985).

\bibitem{BelleUB}
  D.~Liventsev {\it et al.} [Belle Collaboration],
  ``Search for heavy neutrinos at Belle,''
  Phys.\ Rev.\ D {\bf 87}, no. 7, 071102 (2013)
  doi:10.1103/PhysRevD.87.071102
  [arXiv:1301.1105 [hep-ex], v3 with Erratum (2017)].

  
  \bibitem{DELPHI}
  P.~Abreu {\it et al.} [DELPHI Collaboration],
  ``Search for neutral heavy leptons produced in Z decays,''
  Z.\ Phys.\ C {\bf 74}, 57 (1997);
  Erratum, Z.\ Phys.\ C {\bf 75}, 580(E) (1997).

\bibitem{LHCbUB}
  R.~Aaij {\it et al.} [LHCb Collaboration],
  ``Searches for Majorana neutrinos in $B^-$ decays,''
  Phys.\ Rev.\ D {\bf 85}, 112004 (2012)
  [arXiv:1201.5600 [hep-ex]];
  ``Search for Majorana neutrinos in $B^- \to \pi^+\mu^-\mu^-$ decays,''
  Phys.\ Rev.\ Lett.\  {\bf 112}, no. 13, 131802 (2014)
  [arXiv:1401.5361 [hep-ex]].

\bibitem{PesShuUB}
  B.~Shuve and M.~E.~Peskin,
  ``Revision of the LHCb limit on Majorana neutrinos,''
  Phys.\ Rev.\ D {\bf 94}, no. 11, 113007 (2016)
  [arXiv:1607.04258 [hep-ph]].
  
\bibitem{IZ}
C.~Itzykson and J.-B.~Zuber, {\it Quantum Field Theory}, McGraw-Hill (1980), 705 pp.

\bibitem{GiSi}
F.~J.~Gilman and R.~L.~Singleton,
``Analysis of semileptonic decays of mesons containing heavy quarks,''
Phys.\ Rev.\ D {\bf 41}, 142 (1990).

\bibitem{RiBu}
  J.~D.~Richman and P.~R.~Burchat,
  ``Leptonic and semileptonic decays of charm and bottom hadrons,''
  Rev.\ Mod.\ Phys.\  {\bf 67}, 893 (1995)
  [hep-ph/9508250].
  
\bibitem{Belle09}
  I.~Adachi {\it et al.} [Belle Collaboration],
  ``Measurement of the form factors of the decay $B^+ \to {\bar D}^{*0} \ell^+ \nu_{\ell}$ and determination of the CKM matrix element $|V_{cb}|$,''
  arXiv:0910.3534 [hep-ex].
  

\bibitem{Belle18}  
  A.~Abdesselam {\it et al.} [Belle Collaboration],
  ``Measurement of CKM Matrix Element $|V_{cb}|$ from $\bar{B}^{0} \to D^{*+} \ell^{-} \bar{\nu}_\ell$,''
  arXiv:1809.03290 [hep-ex].

\end{thebibliography}
\end{document}